\documentclass[11pt]{article}

\usepackage[final]{acl}

\usepackage{tabularx}
\usepackage{amsmath}
\usepackage{enumitem}
\usepackage{makecell}
\usepackage{booktabs}
\usepackage{siunitx}
\usepackage{times}
\usepackage{latexsym}
\usepackage{adjustbox}

\usepackage{amssymb}
\usepackage[table]{xcolor}
\usepackage{pifont}   
\usepackage{graphicx}
\usepackage{ragged2e}
\usepackage{colortbl}

\usepackage{listings}
\usepackage{textcomp}
\usepackage[most]{tcolorbox}
\usepackage{array}
\usepackage{caption}
\usepackage{subcaption}
\usepackage{placeins}
\usepackage{algorithm}
\usepackage{algpseudocode}

\lstdefinestyle{edme}{
  language=Python,
  basicstyle=\ttfamily\footnotesize,
  columns=fullflexible,
  keepspaces=true,
  upquote=true,
  showstringspaces=false,
  breaklines=true,
  breakatwhitespace=true,
  linewidth=\columnwidth,
  tabsize=2,
  aboveskip=4pt,
  belowskip=4pt,
  xleftmargin=0.35em,
  frame=single,
  framerule=0.2pt,
  rulecolor=\color{black!20},
  backgroundcolor=\color{black!1},
  keywordstyle=\color{blue!55!black},
  commentstyle=\color{green!40!black},
  stringstyle=\color{orange!45!black},
}
\lstdefinestyle{dvp}{
  language=Python,
  basicstyle=\ttfamily\footnotesize,
  columns=fullflexible,
  keepspaces=true,
  upquote=true,
  showstringspaces=false,
  breaklines=true,
  breakatwhitespace=true,
  linewidth=\columnwidth,
  tabsize=2,
  aboveskip=4pt,
  belowskip=4pt,
  xleftmargin=0.35em,
  frame=single,
  framerule=0.2pt,
  rulecolor=\color{black!20},
  backgroundcolor=\color{black!1},
  keywordstyle=\color{blue!55!black},
  commentstyle=\color{green!40!black},
  stringstyle=\color{orange!45!black},
}
\lstdefinestyle{tracefig}{
  style=edme,
  basicstyle=\ttfamily\scriptsize,
  linewidth=\linewidth,
  aboveskip=3pt,
  belowskip=3pt,
  frame=single,
  xleftmargin=0pt,
  xrightmargin=0pt
}
\algrenewcommand{\algorithmicrequire}{\textbf{Input:}}
\algrenewcommand{\algorithmicensure}{\textbf{Output:}}
\algrenewcommand{\algorithmiccomment}[1]{\hfill{\color{black!60}$\triangleright$}~#1}
\algrenewcommand{\alglinenumber}[1]{\scriptsize #1}
\newcommand{\algstage}[1]{\Statex{\color{black!70}\textbf{#1}}}

\newcommand{\cmark}{{\color{green!60!black}\ensuremath{\checkmark}}} 
\newcommand{\xmark}{{\color{red!80!black}\ensuremath{\times}}}            
\newcommand{\appsec}[1]{\vspace{2pt}\noindent\textbf{#1}\hspace{6pt}}

\newtcolorbox{edmebox}[1][]
{colback=blue!2!white,
 colframe=blue!45!black,
 coltitle=blue!45!black,
 fonttitle=\bfseries,
 boxrule=0.4pt,
 arc=2pt,
 left=6pt,
 right=6pt,
 top=6pt,
 bottom=6pt,
 enhanced,
 breakable,
 title=#1,
 before skip=6pt,
 after skip=6pt}

\newtcolorbox{dvpbox}[1][]
{colback=violet!3!white,
 colframe=violet!50!black,
 coltitle=violet!50!black,
 fonttitle=\bfseries,
 boxrule=0.4pt,
 arc=2pt,
 left=6pt,
 right=6pt,
 top=6pt,
 bottom=6pt,
 enhanced,
 breakable,
 title=#1,
 before skip=6pt,
 after skip=6pt}

\newtcolorbox{agentbox}[1][]
{colback=green!3!white,       
 colframe=green!35!black,    
 coltitle=white,
 fonttitle=\bfseries,
 boxrule=0.4pt,
 arc=2pt,
 left=6pt,
 right=6pt,
 top=6pt,
 bottom=6pt,
 enhanced,
 breakable,
 title=#1,
 before skip=6pt,
 after skip=6pt}

\usepackage[T1]{fontenc}

\usepackage[utf8]{inputenc}

\usepackage{microtype}

\usepackage{inconsolata}

\usepackage{graphicx}

\author{Jia Li \\
  The Chinese University of Hong Kong \\
  Hong Kong, Hong Kong \\
  \texttt{linsayli@link.cuhk.edu.hk} \\
  \And
  Yuxin Su\thanks{Corresponding author.} \\
  Sun Yat-sen University \\
  Zhu Hai, China \\
  \texttt{suyx35@mail.sysu.edu.cn} \\
  \AND
  Michael R. Lyu \\
  The Chinese University of Hong Kong \\
  Hong Kong, Hong Kong \\
  \texttt{lyu@cse.cuhk.edu.hk} \\}

\begin{document}
  \title{From Laboratory to Real-World Applications: Benchmarking Agentic Code Reasoning at the Repository Level}
  \maketitle

  \begin{abstract}
As large language models (LLMs) evolve into autonomous agents, evaluating \textit{Repository-Level Reasoning}, the ability to maintain logical consistency across massive, interdependent file systems, has become critical. Current benchmarks typically fluctuate between isolated code snippets and black-box evaluations. We present \textsc{RepoReason}, a white-box diagnostic benchmark centered on \textit{Abductive Assertion Verification}. To eliminate memorization while preserving authentic logical depth, we implement an \textit{Execution-Driven Mutation} framework that utilizes the environment as a \textit{Semantic Oracle} to regenerate ground-truth states. Furthermore, we establish a fine-grained diagnostic system using dynamic program slicing, quantifying reasoning via three orthogonal metrics: \textit{ESV} (Reading Load), \textit{MCL} (Simulation Depth), and \textit{DFI} (Integration Width). Comprehensive evaluations of frontier models (e.g., Claude-4.5-Sonnet, DeepSeek-v3.1-Terminus) reveal a prevalent \textit{Aggregation Deficit}, where integration width serves as the primary cognitive bottleneck. Our findings provide granular white-box insights for optimizing the next generation of agentic software engineering.
\end{abstract}

  \section{Introduction}

The paradigm of software engineering is undergoing a fundamental shift: from human-centric coding to agentic autonomous development \citep{jiang2025agenticsoftwareissueresolution,hou2023large}. As Large Language Models ($\text{LLMs}$) evolve into autonomous software agents, the core competency required has escalated from performing localized execution reasoning within isolated code snippets---akin to controlled laboratory exercises---to maintaining logical consistency across extensive, interdependent file systems---a capability we define as Repository-Level Reasoning. This capability imposes profound challenges on $\text{LLMs}$, demanding not only code synthesis but also the agency to explore vast contexts to pinpoint target definitions, aggregate multi-source information across complex dependencies, and perform long-chain reasoning to accurately track and compute state mutations.

However, a critical gap exists between this growing demand and current evaluation methods. Existing benchmarks largely fall into two extremes. On one side, function-level reasoning datasets, such as $\text{CRUXEval}$ \citep{Gu2024CRUXEval} and $\text{REval}$ \citep{Chen2024REval}, study code execution reasoning within isolated snippets. While these provide analytical rigor, they operate in a laboratory setting that lacks the cross-file dependencies and complex architectural contexts found in real-world software engineering. On the other side, frameworks like $\text{SWE-bench}$ \citep{Jimenez2024SWEbench} evaluate agents in wild repository environments but function as black-boxes: they check whether a problem is solved but offer little insight into why an agent fails. Without detailed diagnosis, the path to improving agent reasoning remains unclear.

To address this gap, we introduce $\text{RepoReason}$, a repository-level code reasoning benchmark designed as a white-box diagnostic tool. Unlike traditional generation tasks, $\text{RepoReason}$ shifts to a verification-centric approach: ``Given the complex execution history of this repository, what is the current system state?'' By requiring models to derive deterministic values that satisfy assertions rather than writing implementation logic, we separate core logical reasoning from syntactic noise (\textit{e.g.}, formatting errors, indentation slips, or API name hallucinations). This ensures that every failure reflects a genuine deficit in reasoning rather than superficial coding errors.

To ensure the benchmark reflects real-world complexity, we extract task instances from a curated set of mainstream Python repositories (\textit{e.g.}, \texttt{toolz} \citep{toolz_repo}, \texttt{sympy} \citep{sympy_repo}, \texttt{jinja2} \citep{jinja2_repo}), challenging models with authentic, deep logic. Beyond static code extraction, we perform concrete execution of the repositories' built-in unit test suites to capture granular runtime traces. By designing structural metrics based on these traces---such as stack depth, function call density, and cross-file dependency breadth---we quantitatively filter and retain only the most complex reasoning tasks. 


Constructing such a benchmark requires navigating between two distinct failure modes of current evaluations: Spurious Correlation (guessing via shallow patterns) and Rote Memorization (recalling via pre-training data). To address this, we propose a two-stage defense mechanism. First, to prevent shortcut learning via visual pattern matching, we reject raw trace values and instead identify Unit Test Assertions as our semantic anchors. While raw traces are often localized and easily inferred from immediate code patterns, assertions function as semantic milestones that encapsulate high-level developer intent, strategically positioned to validate the system state after complex logical operations. They force Abductive Reasoning \citep{josephson1994abductive}, requiring the model to reconstruct the preceding execution history across complex call stacks to deduce the verified state.

However, a significant obstacle remains: many assertions within these mainstream repositories' unit tests have already been encountered by $\text{LLMs}$ during pre-training. This leads to a risk of data leakage \citep{deng2024unveilingspectrumdatacontamination}. To resolve this, we implement an Execution-Driven Mutation methodology. We treat the code execution environment as a Semantic Oracle. By keeping the reasoning logic (the call graph filtered via structural metrics) intact but perturbing the program inputs, we effectively sever the model's memory retrieval path. We then re-execute the entire repository context to capture the new ground-truth state, subsequently masking these verified values to create cloze-style reasoning tasks. Finally, to ensure these captured states are objectively evaluable, we introduce a Deterministic Value Protocol, which filters runtime data to eliminate representational drift and ensure stable, unique ground-truth values. This design ensures that while the reasoning logic remains authentic and complex, the specific state values are unmemorized and deterministic.

Finally, through an extensive evaluation of a diverse range of frontier models including Claude-Sonnet-4.5 and DeepSeek-v3.1-Terminus, we demonstrate that even top-tier agents face severe performance degradation when confronted with high-entropy logic. Our results unveil a critical Aggregation Deficit where accuracy drops sharply as integration width increases, as well as a Cliff Effect in reading comprehension beyond 600 lines and a significant loss of consistency beyond 100 execution steps. These insights highlight that next-generation agents must transcend local reasoning and focus on high-breadth information synthesis and long-chain state consistency.

In summary, this paper makes the following contributions:

\begin{itemize}[noitemsep, leftmargin=4mm]
    \item \textbf{From Prediction to Abductive Verification:} We pioneer a repository-level reasoning paradigm that pivots from direct outcome prediction to abductive assertion verification. By centering tasks on masked assertions within interdependent file systems, we successfully decouple core architectural intelligence from surface-level syntactic noise.
    
    \item \textbf{The Semantic Oracle Framework:} We resolve the Richness-Contamination Dilemma through an Execution-Driven Mutation framework. By utilizing the execution environment as a Semantic Oracle, we sever data leakage paths via input perturbation and real-time state regeneration, ensuring tasks remain logically authentic yet unmemorized.
    
    \item \textbf{White-box Diagnostic Instrument:} Beyond end-to-end scoring, we introduce a granular diagnostic system powered by dynamic program slicing. We formalize three orthogonal cognitive metrics (Reading Load, $\text{ESV}$; Simulation Depth, $\text{MCL}$; Integration Width, $\text{DFI}$) that map reasoning failures to specific cognitive bottlenecks: context overload, state tracking deficits, and aggregation barriers.

    \item \textbf{Empirical Unveiling of Aggregation Deficits:} Through extensive evaluation of frontier models, we provide the first quantitative analysis of agentic reasoning trajectories across massive codebases (up to 776k+ LoC). Our findings identify $\text{DFI}$ as the primary bottleneck for next-generation agents, empirically confirming a severe aggregation deficit in information synthesis.
\end{itemize}
  \begin{table*}[t]
\centering
\renewcommand{\arraystretch}{1.3}
\setlength{\tabcolsep}{5pt}

\caption{Comparison of \textbf{RepoReason} with existing code reasoning and understanding benchmarks.}
\label{tab:comparison}

\resizebox{\textwidth}{!}{
\begin{tabular}{l c c l l l}
\toprule
\textbf{Benchmark} & \textbf{Repo-Lvl} & \textbf{Scale (LoC)} & \textbf{Evaluation Paradigm} & \textbf{Metric Granularity} & \textbf{Diagnostic Depth} \\ \midrule
\rowcolor[HTML]{F9F9F9}
HumanEval \citep{Chen2021HumanEval}    & \xmark & $\sim$10 & Code Generation & Pass/Fail & Black-box \\
MBPP \citep{Austin2021MBPP}            & \xmark & $\sim$10 & Code Generation & Pass/Fail & Black-box \\
\rowcolor[HTML]{F9F9F9}
ClassEval \citep{Du2023ClassEval}      & \xmark & $\sim$40 & Code Generation & Pass/Fail & Black-box \\
CRUXEval \citep{Gu2024CRUXEval}        & \xmark & $\sim$10 & Execution Prediction & Pass/Fail & Black-box \\
\rowcolor[HTML]{F9F9F9}
REval \citep{Chen2024REval}            & \xmark & $\sim$10--40 & Runtime Behavior & Pass/Fail & White-box (Local) \\
CORE \citep{Xie2025CORE}               & \xmark & 21--100 & Static Analysis & Pass/Fail & White-box (Static) \\
\rowcolor[HTML]{F9F9F9}
SWE-bench \citep{Jimenez2024SWEbench}  & \cmark & 1k -- 1M+ & Issue Resolution & Pass/Fail & Black-box \\ \midrule
\rowcolor[HTML]{E8E8E8}
\textbf{RepoReason (Ours)} & \cmark & \textbf{1.2k -- 775k+} & \textbf{Assertion Verification} & \textbf{Cognitive Metrics} & \textbf{White-box (Global)} \\ \bottomrule
\end{tabular}
}
\end{table*}

\section{Related Work}

\subsection{Code Generation}
Existing code benchmarks have evolved from function-level synthesis (\textit{e.g.}, HumanEval \citep{Chen2021HumanEval}, MBPP \citep{Austin2021MBPP}) to class-level and repository-level completion (\textit{e.g.}, ClassEval \citep{Du2023ClassEval}, RepoBench \citep{Liu2023RepoBench}). While these frameworks increasingly reflect real-world complexity, they primarily target syntactic completion---the ability to produce correct code text. In contrast, $\text{RepoReason}$ shifts the focus to semantic reasoning. By requiring the deduction of deterministic system states via assertion verification, it isolates core logical reasoning from the syntactic noise and hallucination risks inherent in code generation.

\subsection{Code Reasoning}
Recent benchmarks like $\text{CRUXEval}$ \citep{Gu2024CRUXEval} and $\text{REval}$ \citep{Chen2024REval} pioneered execution reasoning but are largely confined to isolated, lab-scale code snippets. While tools like $\text{CORE}$ \citep{Xie2025CORE} incorporate static analysis tasks to evaluate reasoning, they lack the structural depth and code breadth (\textit{e.g.}, deep inheritance, cross-module imports) characteristic of large-scale projects. $\text{RepoReason}$ bridges this gap by applying deep execution simulation to massive repository scopes, challenging models to maintain logical consistency across extensive file systems.

\subsection{Agentic Software Engineering}
At the highest level, $\text{SWE-bench}$ \citep{Jimenez2024SWEbench} evaluates agents on resolving GitHub issues. However, as an end-to-end evaluation, it acts as a black-box indicator of success or failure. $\text{RepoReason}$ complements this by targeting the cognitive root causes of failure. By decoupling reasoning from code editing and utilizing fine-grained metrics ($\text{ESV}$, $\text{MCL}$, $\text{DFI}$), it provides the white-box diagnostic insights required to pinpoint an agent's specific bottlenecks in context handling or multi-source integration.

\subsection{Summary Comparison}
Table~\ref{tab:comparison} provides a multi-dimensional comparison between $\text{RepoReason}$ and mainstream code benchmarks, highlighting its unique advantages in diagnostic depth and structural complexity.
  \begin{figure*}
    \centering
    \includegraphics[width=\textwidth]{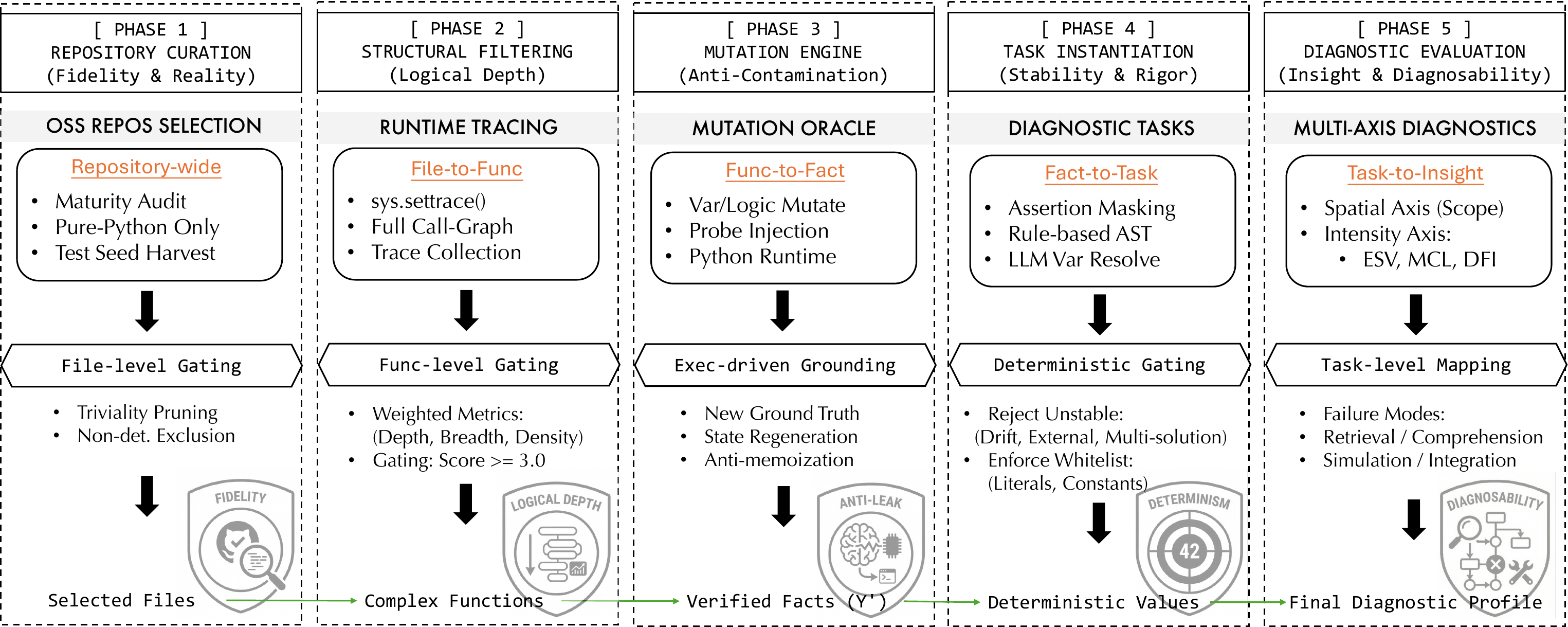}
    \caption{The overall architecture and multi-stage pipeline of RepoReason, encompassing repository curation, structural filtering, execution-driven mutation, task instantiation, and diagnostic evaluation.}
    \label{fig:arch}
\end{figure*}

\section{RepoReason: Design and Construction}
\label{sec:design}

In this section, we delineate the methodological framework of $\text{RepoReason}$, conceptualizing the construction pipeline as a rigorous five-phase defense system. Our objective is to transform authentic software logic into white-box diagnostic tasks while systematically neutralizing systemic vulnerabilities: spurious correlation, rote memorization, and logical superficiality. As illustrated in Figure~\ref{fig:arch}, the pipeline orchestrates sequential gates: (1) Repository Curation to ensure environmental fidelity; (2) Structural Filtering via runtime traces to guarantee logical depth; (3) Execution-Driven Mutation to eliminate parametric leakage; (4) Task Instantiation governed by a deterministic value protocol for stability; and (5) Diagnostic Evaluation establishing fine-grained cognitive profiles. This integration ensures that each task instance necessitates authentic repository-level reasoning and robust state tracking, effectively precluding resolution via shallow pattern matching. 

\subsection{Design Philosophy: Assertions as Semantic Anchors}

We explicitly reject two prevailing paradigms: traditional Input/Output (I/O) Prediction and Raw Trace Querying. 

\textbf{I/O Prediction Inefficiency}: The traditional I/O Prediction paradigm \citep{Gu2024CRUXEval,Chen2024REval}, while effective for isolated algorithmic puzzles, is fundamentally ill-suited for repository-level reasoning. In authentic software ecosystems, runtime data structures, such as SymPy's symbolic expression trees or NetworkX's graph topologies, are often voluminous and complex, lacking the clear, concise boundaries found in synthetic tasks. Consequently, verifying the entire output state is inherently inefficient and noise-laden, as a significant portion of the data consists of dynamic metadata or irrelevant state rather than the target logic. 

\textbf{Spurious Correlation in Raw Traces}: A straightforward approach might involve randomly querying variable values at specific locations (e.g., ``What is the value of variable $x$ at line 10?''). However, this paradigm of raw trace querying is fundamentally flawed for evaluating deep reasoning. It suffers from shortcut learning, as models can often retrieve answers through shallow static analysis of nearby assignment statements, bypassing the need for genuine execution simulation. 

\textbf{The Solution: Assertions as Semantic Anchors}. To overcome these pitfalls, we identify unit test assertions as the optimal semantic units for instantiation. Assertions are not random data points; they are crystallized checkpoints of developer intent, strategically placed to verify critical state transitions after complex operations. By targeting assertions, we filter out implementation noise and focus solely on the logical invariants of the system.
We employ a Mask-and-Reason strategy: we take a valid assertion (e.g., \texttt{assert len(cache) == 5}) and mask the ground truth value (e.g., \texttt{assert len(cache) == <mask>}). This transforms the task into Abductive Reasoning: the model cannot find the answer by reading the immediate text. Instead, it must mentally reconstruct the preceding execution history, tracing data flows across files and simulating state mutations, to deduce the only logical value that satisfies the verification condition.

\subsection{Foundation: Real-world Repository Curation}

To ensure that our semantic anchors are grounded in reality, we curated a set of mature, actively maintained Python repositories. This stage ensures the fidelity of our benchmark, providing the complex, cross-file dependency structures that synthetic datasets lack.

\subsubsection{Repository Selection}

We selected mature, actively maintained, and predominantly pure Python open-source projects from $\text{PyPI}$ and $\text{GitHub}$. Our selection specifically targets libraries where the core logic is implemented in Python, enabling complete line-level tracing via \texttt{sys.settrace}. Libraries with substantial C/C++/Cython extensions (\textit{e.g.}, NumPy, Pandas, PyTorch, scikit-learn) were explicitly excluded, as their core computational routines remain opaque to Python-level instrumentation.

The selected repositories cover diverse reasoning domains and exhibit diverse complexity, with codebases ranging from approximately 1,200 to over 775,000 lines of code (LoC). This deliberate variation allows for a comprehensive evaluation of LLMs' code reasoning capabilities across different scales and perspectives. Detailed description and statistics are shown in Table~\ref{tab:repo_stats}, Section~\ref{sec:repository}.

\subsubsection{File-Level Semantic Filtering}

Test selection prioritizes cases requiring multi-step semantic reasoning over superficial pattern matching. We manually review source code at the file level to screen all test modules. This initial stage aims to filter out low-value assertions and ensure Deterministic Verifiability. We systematically excluded test files falling into three specific categories: (1) Trivial API Validation, comprising simple attribute access (\textit{e.g.}, \texttt{assert url.scheme == 'http'}), direct dictionary operations without cross-module dependencies (\textit{e.g.}, \texttt{assert d.get('key') == 'value'}), and metadata introspection lacking computational reasoning (\textit{e.g.}, \texttt{assert num\_required\_args(lambda x: None) == 1}); (2) Shallow Computation Patterns, including direct formula application or lookup tables (\textit{e.g.}, \texttt{assert quote('a b') == 'a\%20b'}), simple string manipulation without state complexity, and local transformations requiring no cross-module reasoning (\textit{e.g.}, \texttt{assert x + 0 == x}); and (3) Non-deterministic or External Dependencies, such as tests relying on randomness (\textit{e.g.}, \texttt{assert random.choice(seq) in seq}) or external timing or system state (\textit{e.g.}, \texttt{assert time.time() > start}).

\subsection{Countering Memorization: The Execution-Driven Mutation Engine} 

The second line of defense addresses Rote Memorization. Since powerful LLMs have likely memorized public codebases, using raw GitHub code leads to data contamination. Our Execution-Driven Mutation framework serves as an Anti-Contamination Engine, creating parallel reasoning realities that preserve logic but alter facts. 

\textbf{Unified Semantic and Logic Mutation.} We employ a unified mutation strategy utilizing a teacher LLM. In a single step, the model performs both visual changes (\textit{e.g.}, renaming variables, restructuring code, and removing comments) and logic changes (\textit{e.g.}, modifying constants, input data, and parameters) \citep{Jia2011Mutation}. This combined approach ensures the mutated code is correct and readable, while effectively removing visual clues and invalidating memorized answers. Crucially, it strictly preserves the original API call sequence to the target library to maintain invariant reasoning complexity. 

\textbf{Execution-Driven Ground Truth Regeneration.} To ensure the correctness of mutated tests and avoid $\text{LLMs}$ hallucinations, we designed an Execution-Driven Assertion Reconstruction framework. First, we use Probe Injection, where the $\text{LLMs}$ converts original assertions into print statements formatted as \texttt{DEBUG\_RESULT}, turning the test into code that outputs its internal state. Next, we run this code in a controlled environment to capture the actual Runtime Values produced by the logic changes. These values serve as the ground truth, derived directly from code execution rather than model guessing. Finally, we perform Style-Consistent Assertion Generation. Using the captured values and the original code's style as guides, the $\text{LLMs}$ reconstructs the assertions. It mimics the original writing patterns, \textit{for example}, keeping \texttt{is} for object identity checks instead of changing to \texttt{==}, or using specific constructors like \texttt{FiniteSet(1, 2)} instead of native sets like \texttt{\{1, 2\}}. This ensures the new ground truth is mathematically accurate and matches the original codebase's style.

\textbf{Strict Validation Gate.} We enforce a strict validation process. A mutated test is considered valid only if it meets three criteria: (1) Executability, where the code runs without errors; (2) Assertion Validity, ensuring new assertions pass against the mutated logic; and (3) API Preservation, where the API call sequence matches the original test. Any variant failing these checks is discarded, ensuring all tasks in $\text{RepoReason}$ are solvable, deterministic, and maintain original reasoning depth.

\subsection{Countering Superficiality: Trace-Based Structural Filtering} 

The third defense targets Logical Superficiality. Even in complex repositories, many tests remain shallow. To ensure RepoReason evaluates genuine Repository-Level capabilities, we employ Trace-Based Structural Filtering (Phase 2) to identify tasks with sufficient logical depth. 

\textbf{Trace Collection.} With the mutated test cases ready, we capture their runtime behavior using a lightweight tracing system based on Python's built-in \texttt{sys.settrace()} function, which enables line-level execution tracking. Our system records the complete sequence of function calls during the execution of the mutated tests, capturing granular details including function names, module paths, line numbers, call order, call stack depth, function source code, input arguments, return values, and executed lines within each function. 
Critically, to eliminate noise, we automatically identify the test function as the entry point and focus strictly on the target library's execution path. Furthermore, to prevent infinite recursion and ensure data manageability, we enforce a maximum call depth of 3, where the test function itself is defined as depth 0.

\textbf{Function-Level Structural Filtering and Complexity Grading.} Instead of traditional hard thresholds, we implemented a weighted scoring system based on captured traces to identify high-density test cases. 
We calculated a complexity score for each test function: 

{
\setlength{\abovedisplayskip}{2pt} 
\setlength{\belowdisplayskip}{2pt} 
\begin{equation}
\small
\begin{aligned}
\text{Score} = & \left(\frac{N_{\text{files}}}{4} \times 0.1\right) + \left(\frac{N_{\text{funcs}}}{15} \times 0.2\right) \\
& + \left(\frac{N_{\text{calls}}}{30} \times 0.5\right) + \left(\frac{D}{4} \times 0.2\right)
\end{aligned}
\label{eq:complexity_score}
\end{equation}
}

This formula prioritizes computational intensity (Call Count, 50\%) and execution depth (Stack Depth, 20\%), while balancing coverage breadth (Function Count, 20\%) and cross-module dependencies (File Count, 10\%). 
The baselines ($N_{\text{files}}=4, N_{\text{funcs}}=15, N_{\text{calls}}=30, D=4$) are calibrated to the empirical distribution of the target repositories. 
We primarily select tests rated 3 stars or higher ($\text{Score} \ge 0.30$), filtering out approximately 13\% of trivial test cases. 

\subsection{Countering Ambiguity: The Deterministic Value Protocol}
\label{sec:deterministic_protocol}

To ensure reproducibility and eliminate evaluation noise, we introduce the Deterministic Value Protocol. This protocol enforces that ground-truth answers are unique and stably representable through two programmatic filtering funnels (covering $\sim$90\% of samples):

\textbf{Semantic Determinism Funnel}: Automatically filters samples with inherent uncertainty, including: (1) representational drift (\textit{e.g.}, strings containing random memory addresses); (2) external perturbations (\textit{e.g.}, non-deterministic signals like timestamps or UUIDs); and (3) multi-solution semantics (\textit{e.g.}, range comparisons like \texttt{assert x > 10}).

\textbf{Morphological Determinism Funnel}: Mandates that masked values represent concrete runtime states rather than arbitrary code text. A whitelist priority is established: Literals $>$ Global Constants $>$ Parameter-Resolvable Constructors. As a special case, \textit{sympy} allows complex mathematical expressions (\textit{e.g.}, \texttt{Rational(33, 2) - Rational(11, 2)*sqrt(3)}) as valid answers, as the expression itself serves as the standard storage form. Pure variable references or dynamic function calls are strictly rejected.

For descriptive variables used by developers (\textit{e.g.}, \texttt{assert result == expected}), the system activates an LLM Variable Resolver to unfold abstract references into runtime-verified literal forms, ensuring evaluation is anchored to deterministic system states.

\subsection{Cognitive Diagnostic Framework}
\label{sec:diagnostic_framework}

To quantify the intrinsic reasoning complexity of a task, we perform dynamic program slicing \citep{ agrawal1990dynamic} rooted at the assertion to obtain the minimal causal computational subgraph along the actual execution path. Dynamic cross-function backward slicing is obtained by tracing back from the assertion along data and control dependencies within a recorded execution trace to isolate the minimal set of statements that causally influence the observed result. Based on this subgraph, we characterize reasoning intensity using three mutually orthogonal metrics rooted in cognitive load theory \citep{sweller1988cognitive}.

\textbf{1. Reading Load ($\text{ESV}$ --- Effective Sliced Volume)}: Measures the causally relevant source code volume required to process. We quantify the source volume of computational units (functions/methods) in the slice rather than simple slice line counts, as reasoning requires restoring semantic context (parameters, constraints, preconditions) within stable boundaries. Failures at high $\text{ESV}$ signify \textit{Comprehension Failure} rooted in \textit{Context Overload}, where the agent loses semantic focus across the extended causal chain.

\textbf{2. Simulation Depth ($\text{MCL}$ --- Mutation Chain Length)}: Measures the number of dynamic state update steps required to reach the target state. We treat each causally relevant statement in the dynamic slice as a state advancement atomic step and sum them weighted by execution frequency. This metric reflects the difficulty of maintaining a virtual machine state. Significant performance deterioration at high $\text{MCL}$ indicates \textit{Simulation Failure} stemming from a \textit{State Tracking Deficit}, where the agent struggles to accurately synchronize variable states through complex sequences, leading to the breakdown of deep simulation.

\textbf{3. Integration Width ($\text{DFI}$ --- Dependency Fan-in)}: Measures the number of independent upstream logical inputs that collectively determine the critical value at an assertion. It quantifies logical inputs consumed by the execution subgraph but generated outside of it, while explicitly filtering out structural noise such as language keywords or module imports. Performance degradation at high $\text{DFI}$ reveals an \textit{Integration Failure} caused by an \textit{Aggregation Deficit} in synthesizing disparate logical sources into a coherent conclusion.

  \begin{figure*}
    \centering
    \includegraphics[width=\textwidth]{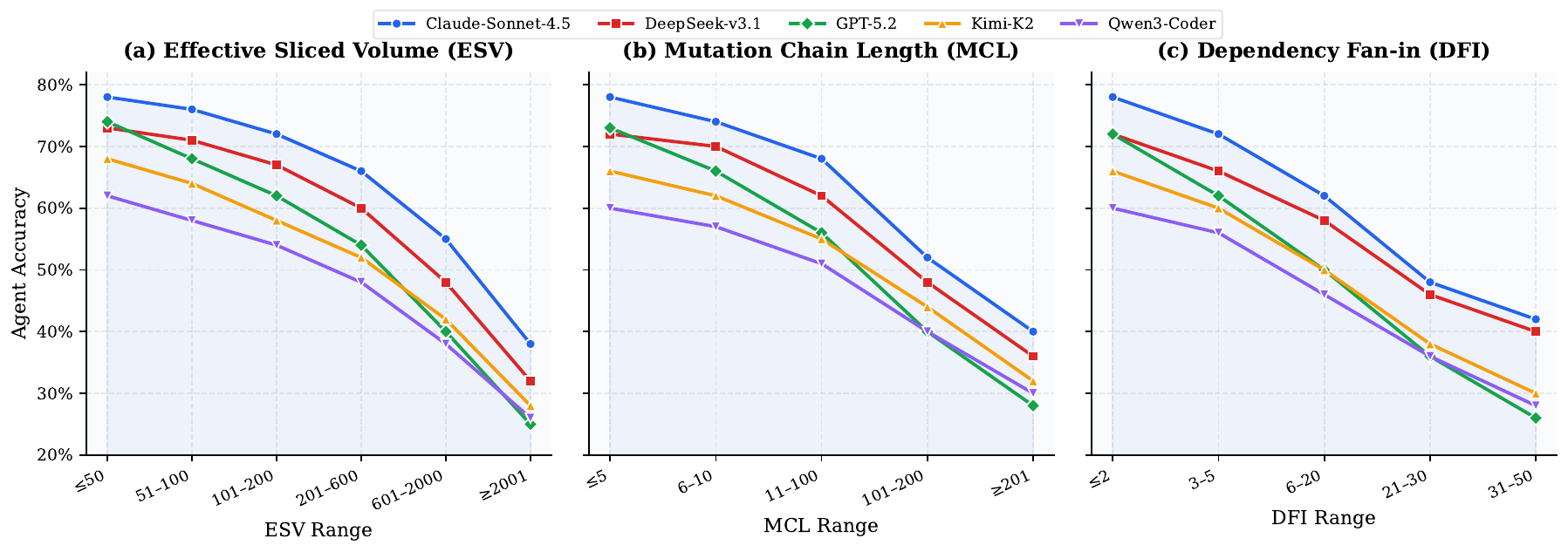}
    \caption{Performance trajectories of frontier LLM agents across three orthogonal cognitive metrics: (a) Effective Sliced Volume (ESV), (b) Mutation Chain Length (MCL), and (c) Dependency Fan-in (DFI).}
    \label{fig:cognitive}
\end{figure*}

\section{Experimental Evaluation}

To validate the diagnostic effectiveness of RepoReason, we conducted a comprehensive evaluation using state-of-the-art LLMs.

\subsection{Experimental Setup}

\textbf{Agent Framework.}
We utilized OpenHands (version 0.60.0) \citep{OpenHands2024} as our evaluation platform. Specifically, we configured the ReadOnlyAgent for all tasks. This agent configuration is restricted to file system exploration and code reading operations, preventing any modification to the repository. This setup perfectly aligns with the verification nature of RepoReason, isolating the model's ability to ``read, navigate, and reason'' from its ability to ``edit or plan.''

\textbf{Model Selection.}
We evaluated a diverse set of frontier models representing the latest advancements in code reasoning: the \textbf{Claude Family} (claude-sonnet-4.5 \citep{Claude45_2025}), the \textbf{GPT Family} (gpt-5.2 \citep{GPT5_2025}), the \textbf{DeepSeek Family} (deepseek-3.1-terminus \citep{DeepSeek2024}), the \textbf{Kimi Family} (Kimi-K2 \citep{KimiK2_2025}), and the \textbf{Qwen Family} (qwen3-coder-480b-a35b \citep{Qwen3_2025}).

\textbf{Inference Configuration.}
To ensure reproducibility and eliminate generation randomness, we set the temperature to 0 for all model evaluations. Additionally, to accommodate the potentially lengthy reasoning chains required for repository-level tasks, we set the maximum generation length (max\_tokens) to 8192 for each inference step.

\subsection{Overall Performance: Accuracy across Repositories and Difficulty Levels}

We report the Pass@1 accuracy of the four model families across the seven selected repositories. To control for task variance, results are stratified into Easy ($N=1009$), Medium ($N=838$), and Hard ($N=645$) groups. Task difficulty is annotated per instance by an independent LLM (GPT-5.2) during the generation phase using a structured prompt based on AI reasoning complexity, independent of the evaluated models:
\textbf{Easy}---basic Python knowledge, direct code reading, minimal state tracking;
\textbf{Medium}---non-trivial API behavior, simple state mutations, control flow loops;
\textbf{Hard}---deep domain knowledge, multi-layer dependencies, extensive state tracking, or metaprogramming.
The complete classification prompt is provided in Appendix~\ref{app:difficulty}. The total sample size is Overall ($N=2492$).

Table~\ref{tab:overall_acc} summarizes the overall performance of the agents. All models exhibit a clear performance degradation from Easy to Hard tasks, with Claude-Sonnet-4.5 maintaining the highest accuracy across all difficulty levels.

\begin{table}[t]
\footnotesize
\centering
\renewcommand{\arraystretch}{1.25}
\caption{Overall Agent Accuracy by Difficulty (\%). \textbf{Bold} = best; \underline{underline} = second.}
\label{tab:overall_acc}
\setlength{\tabcolsep}{4pt}
\begin{tabularx}{\columnwidth}{@{} l *{4}{>{\centering\arraybackslash}X} @{}}
\toprule
\textbf{Model} & \textbf{Overall} & \textbf{Easy} & \textbf{Medium} & \textbf{Hard} \\
\midrule
\rowcolor[HTML]{F5F5F5}
Claude-Sonnet-4.5      & \textbf{66.98} & \textbf{80.27} & \textbf{66.79} & \textbf{47.39} \\
DeepSeek-v3.1-Terminus & \underline{60.96} & 74.03 & \underline{60.02} & \underline{41.71} \\
\rowcolor[HTML]{F5F5F5}
GPT-5.2                & 56.86 & \underline{75.92} & 53.94 & 30.85 \\
Kimi-K2                & 54.74 & 69.38 & 51.79 & 35.66 \\
\rowcolor[HTML]{F5F5F5}
Qwen3-Coder-480B       & 50.56 & 61.35 & 49.16 & 35.50 \\
\bottomrule
\end{tabularx}
\end{table}

Through a detailed analysis of Table~\ref{tab:repo_breakdown}, we observe three distinct categories:

\textbf{1. Explicit Algorithmic Patterns (Cachetools and Toolz)}: In repositories such as \texttt{cachetools} and \texttt{toolz}, models demonstrate robust and consistent performance, with mean accuracies ranging from 78.00\% to 82.00\%. These libraries primarily rely on standard algorithmic paradigms---such as cache eviction and functional data flows---characterized by high logical locality and linear execution transitions. The results indicate that $\text{LLMs}$ have effectively mastered these procedural patterns, which currently define the cognitive comfort zone for agentic reasoning.

\textbf{2. Implicit Metaprogramming Logic (Attrs and Jinja2)}: \texttt{attrs} and \texttt{jinja2} exhibit a significant performance stratification. Taking \texttt{attrs} as an example, which relies heavily on decorators for implicit class generation (\textit{e.g.}, the dynamic injection of \texttt{\_\_init\_\_} methods), the performance gap between Claude (72.41\%) and Kimi-K2 (43.10\%) reveals a fundamental divergence in the internalization of the Python dynamic object model. Frontier models demonstrate the capacity to model runtime behaviors that remain invisible as explicit code lines in static text, whereas other models show a marked deficiency when logic is generated programmatically rather than defined literally.

\textbf{3. High-Entropy Symbolic and Structural Reasoning (SymPy and NetworkX)}: \texttt{sympy} and \texttt{networkx} represent the primary reasoning ceiling for current agents. Symbolic transformations in \texttt{sympy} involve abstract, non-linear logical transitions that frequently cause reasoning chains to deviate. At the same time, \texttt{networkx} requires high structured working memory to track complex graph topologies. Claude's superior performance in \texttt{networkx} (79.77\%) confirms its advanced ability to maintain accurate state snapshots during complex traversals without experiencing memory decay or structural confusion.

\begin{table}[htbp]
\scriptsize 
\centering
\renewcommand{\arraystretch}{1.35} 
\caption{Agent Accuracy Breakdown by Repository (\%). \textbf{Bold} = best; \underline{underline} = second.}
\label{tab:repo_breakdown}
\setlength{\tabcolsep}{1.8pt} 
\begin{tabularx}{\columnwidth}{@{} l *{7}{>{\centering\arraybackslash}X} @{}}
\toprule
\textbf{Model} & \makecell{\textbf{cache} \\ \scriptsize(193)} & \makecell{\textbf{toolz} \\ \scriptsize(320)} & \makecell{\textbf{yarl} \\ \scriptsize(192)} & \makecell{\textbf{netx} \\ \scriptsize(404)} & \makecell{\textbf{jinja} \\ \scriptsize(325)} & \makecell{\textbf{attrs} \\ \scriptsize(58)} & \makecell{\textbf{symp} \\ \scriptsize(1000)} \\
\midrule
\rowcolor[HTML]{F5F5F5}
Claude-4.5   & \textbf{79.79} & \textbf{82.81} & \underline{78.65} & \textbf{79.77} & 67.46 & \textbf{72.41} & \textbf{51.10} \\
Deep-3.1     & 78.24 & \underline{77.19} & 70.83 & 63.37 & \textbf{74.15} & \underline{62.07} & \underline{45.20} \\
\rowcolor[HTML]{F5F5F5}
GPT-5.2      & \underline{78.76} & 72.19 & \textbf{81.77} & \underline{65.10} & \underline{66.46} & 65.52 & 36.00 \\
Kimi-K2      & \underline{78.76} & 62.81 & 52.08 & 70.54 & 60.92 & 43.10 & 39.60 \\
\rowcolor[HTML]{F5F5F5}
Qwen3-Coder  & 70.98 & 62.81 & 59.90 & 57.43 & 55.38 & 43.10 & 37.00 \\
\bottomrule
\end{tabularx}
\end{table}

\subsection{Diagnostic Analysis via Cognitive Metrics}

To provide a systematic white-box diagnosis of agentic code reasoning, we analyze model performance across three orthogonal cognitive dimensions: Reading Load ($\text{ESV}$), Simulation Depth ($\text{MCL}$), and Integration Width ($\text{DFI}$).

Figure~\ref{fig:cognitive} visualizes the performance trajectories of all model families as a function of the three cognitive intensity metrics. We observe a consistent and pronounced inverse relationship across all axes: accuracy systematically degrades as $\text{ESV}$, $\text{MCL}$, and $\text{DFI}$ increase. This trend empirically validates our diagnostic framework, demonstrating that $\text{RepoReason}$ effectively quantifies the escalation of cognitive load in repository-level reasoning and serves as a robust discriminator for architectural intelligence.

A comparative analysis of the decline slopes identifies $\text{DFI}$ as the primary cognitive bottleneck for next-generation software agents. The slope of performance degradation is steepest along the $\text{DFI}$ axis across all models. When $\text{DFI}$ exceeds 20 independent upstream sources, accuracy for most models, with the exception of Claude, falls below the 40\% threshold, revealing an Aggregation Deficit: agents struggle to parallelly hold and synthesize constraints from multiple disparate logical sources.

Regarding context processing, we identify a clear ``Cliff Effect'' in $\text{ESV}$. Performance remains relatively stable in lower context ranges but exhibits a sharp decline once the volume of causally relevant computational units exceeds 600 LoC. This threshold marks the transition into Context Overload, where agents fail to maintain semantic focus across the sliced causal chain. Similarly, for $\text{MCL}$, we observe a significant deterioration in logical consistency beyond 100 execution steps. 

Table~\ref{tab:correlation_metrics} provides the precise Pearson correlation coefficients for these trends. The statistical evidence confirms that while reading load and update depth are significant factors, $\text{DFI}$ exhibits the strongest negative correlation with accuracy, reaching $-$0.234 for GPT-5.2. This confirms that Integration Failure is the most severe obstacle in repository-level reasoning, necessitating that future research prioritize enhancing agents' capabilities in high-breadth information synthesis.

\begin{table}[htbp]
\footnotesize
\centering
\renewcommand{\arraystretch}{1.25}
\caption{Pearson Correlation between Cognitive Metrics and Accuracy. \textbf{Bold} = strongest per row.}
\label{tab:correlation_metrics}
\setlength{\tabcolsep}{4pt}
\begin{tabularx}{\columnwidth}{@{} l *{3}{>{\centering\arraybackslash}X} @{}}
\toprule
\textbf{Model} & \textbf{ESV} & \textbf{MCL} & \textbf{DFI} \\
\midrule
\rowcolor[HTML]{F5F5F5}
GPT-5.2                & $-$0.188 & $-$0.158 & \textbf{$-$0.234} \\
Claude-Sonnet-4.5      & $-$0.161 & $-$0.122 & \textbf{$-$0.225} \\
\rowcolor[HTML]{F5F5F5}
DeepSeek-v3.1-Terminus & $-$0.152 & $-$0.122 & \textbf{$-$0.157} \\
Kimi-K2                & $-$0.130 & $-$0.117 & \textbf{$-$0.195} \\
\rowcolor[HTML]{F5F5F5}
Qwen3-Coder-480B       & $-$0.148 & $-$0.154 & \textbf{$-$0.196} \\
\bottomrule
\end{tabularx}
\end{table}

Analysis of Table~\ref{tab:correlation_metrics} reveals distinct cognitive profiles for each model family, highlighting unique architectural trade-offs. \textit{DeepSeek-v3.1} emerges as the most architecturally balanced model, showing exceptional robustness against high Integration Width (DFI, $-0.157$), whereas \textit{Claude-4.5} excels in simulation depth but remains fragile in multi-source integration ($-0.225$). Furthermore, while \textit{Qwen3-Coder} demonstrates superior resilience to context volume (ESV, $-0.148$) and performs optimally in handling large-scale contexts, its state-tracking consistency lags behind the leaders, and \textit{GPT-5.2} remains globally the most vulnerable model across all dimensions.

While the three metrics exhibit conceptual orthogonality, they correlate numerically in real-world codebases. To isolate the primary bottleneck, we conducted a partial correlation analysis across all models. Even after controlling for ESV and MCL, DFI maintains a highly significant partial correlation with accuracy ($r = -0.127, p < 10^{-9}$), whereas the partial correlations for ESV and MCL become statistically insignificant. This confirms DFI as the true primary barrier to repository-level reasoning.

To concretize this finding, we examine two representative high-DFI failures in detail (full cases in Appendix~\ref{app:dfi_cases}). In Jinja2's \texttt{test\_groupby\_default} (DFI=16), the agent correctly inferred the \texttt{default} parameter semantics but failed to integrate the implicit sorting step of \texttt{groupby}, producing two disjoint groups instead of one merged group. In NetworkX's \texttt{test\_all\_simple\_paths} (DFI=10), the agent accurately constructed the graph topology and enumerated paths but completely ignored the \texttt{cutoff=2} constraint, returning paths with 3--4 edges. Both cases exhibit a common \textit{partial integration followed by breakage} pattern: agents successfully process initial information layers but systematically drop final constraints as the number of independent sources grows. This mechanistic evidence corroborates the statistical finding that DFI is the primary cognitive bottleneck.

\subsection{Ablation Study}

To quantify the impact of our benchmark design choices, we conduct two ablation studies on the DeepSeek-v3.1-Terminus agent. First, we evaluate the Execution-Driven Mutation Engine (EDME). Without EDME (evaluating on unmutated, original repository code), the agent achieves 73.15\% accuracy, which drops significantly to 60.96\% with EDME enabled. This exposes a Pairwise Memorization Rate ($\mathrm{MR} = \Pr(\text{pass}_{\text{orig}} \wedge \neg \text{pass}_{\text{mut}})$) of 10.71\%, validating EDME's effectiveness in mitigating data contamination. Second, we assess the Structural Filtering phase. By comparing unfiltered assertions against our filtered subset, we observe substantial increases in average task complexity across all metrics: mean ESV rises from 327.6 to 393.6 (+20.1\%), mean MCL from 85.3 to 93.9 (+10.1\%), and mean DFI from 6.0 to 8.3 (+38.3\%). This confirms that our filtering mechanism successfully isolates the most cognitively demanding reasoning tasks.

  \section{Conclusion}

We presented \textsc{RepoReason}, a white-box diagnostic benchmark for evaluating repository-level code reasoning in LLM agents. By centering on Abductive Assertion Verification and leveraging an Execution-Driven Mutation framework that treats the runtime environment as a Semantic Oracle, we ensure that tasks are both logically authentic and resistant to data contamination. Our fine-grained diagnostic system, built on dynamic program slicing, quantifies reasoning complexity through three orthogonal cognitive metrics: Reading Load (ESV), Simulation Depth (MCL), and Integration Width (DFI). Comprehensive evaluations of frontier models reveal a prevalent Aggregation Deficit, where integration width serves as the primary cognitive bottleneck, alongside a Cliff Effect in reading comprehension beyond 600 lines and significant consistency loss beyond 100 execution steps. These findings provide actionable insights for the next generation of agentic software engineering systems.

\section{Limitations}

Despite the rigorous design of RepoReason, we identify several areas for future refinement:

\textbf{Language Scope:} Currently, RepoReason focuses primarily on Python. Although Python remains the de facto standard for AI-native development and LLM research, extending this benchmark to languages such as Java, C++, and Rust is a priority for future work. We address this potential limitation by ensuring that our system architecture is built on standard profiling interfaces, which are conceptually portable to other language ecosystems, thereby enabling our white-box diagnostic paradigm to cover a broader range of programming environments.

\textbf{Logical Depth Relative to Base Tests:} The reasoning depth of the generated tasks is partially limited by the complexity of the original repository's test suites. We ensure foundation quality by curating mature projects with exceptionally high test coverage and complexity. The mutation engine then evolves these human-written fixed tests into vast permutations of reasoning scenarios, significantly enhancing task diversity. Incorporating adversarial test generation techniques~\citep{shi2026codehacker} to augment the source test pool is a promising future direction.

\textbf{Trace Depth Optimization:} We employ a maximum call depth limit (\texttt{max\_call\_depth=3}) during dynamic trace extraction. Our analysis reveals that removing this limit inflates trace volume by 6.2x overall, predominantly introducing low-level infrastructure functions (\textit{e.g.}, AST traversers, magic methods) that contribute minimally to semantic understanding. Thus, the depth limit serves as a crucial signal-to-noise ratio optimization rather than an artificial restriction on reasoning capacity.

\textbf{LLM Participation in Pipeline:} While our data construction pipeline utilizes LLMs, their involvement is strictly constrained to two of the five stages (mutation generation and fallback variable resolution, covering $\sim$10\% of samples). Crucially, all ground-truth state values are captured through concrete code execution, not LLM generation. This hybrid approach ensures that potential model biases do not compromise the objective validity of the evaluation targets.

\section{Ethics Statement}
RepoReason is constructed entirely from public repositories with open-source licenses that permit research and software usage.  During data collection and evaluation, we do not collect metadata regarding repository pull request authors or individual contributors, and our pipeline utilizes only publicly available information accessible via standard APIs. Our work did not involve human subject participation; we did not crowdsource or recruit human workers for any stage of the benchmark construction. All manual validation of task instances, environment setup, and metric calibration were conducted solely by the authors.

  \section*{Acknowledgments}
  This work was supported by the Research Grants Council of the Hong Kong Special Administrative Region, China (No. SRFS2425-4S03 of the Senior Research Fellow Scheme and No. CUHK 14209124 of the General Research Fund).

  \bibliography{main}

  \appendix
  \section{Repository Selection and Statistical Overview} \label{sec:repository}
Table~\ref{tab:repo_stats} presents the detailed statistics and core reasoning scenarios for our selected repositories, where lines of code (LoC) are measured using the \texttt{cloc} tool and specifically represent source code by excluding configuration files, documentation, and metadata.

\begin{table}[ht]
\footnotesize
\centering
\caption{Statistics and Key Reasoning Scenarios of Selected Repositories}
\label{tab:repo_stats}
\setlength{\tabcolsep}{3pt}
\begin{tabularx}{\columnwidth}{@{} l r r X @{}}
\toprule
\textbf{Repository} & \textbf{LoC} & \textbf{Tests} & \textbf{Core Reasoning Complexity} \\
\midrule
\makecell[l]{\textbf{cachetools} \\ \cite{cachetools_repo}} & 1.2k & 88 & Cache eviction (LRU/LFU/TTL), state management \\ \addlinespace[2pt]
\makecell[l]{\textbf{yarl} \\ \cite{yarl_repo}}       & 2.3k & 369 & URL parsing, normalization, codec consistency \\ \addlinespace[2pt]
\makecell[l]{\textbf{toolz} \\ \cite{toolz_repo}}     & 3.7k & 92 & Func. flow, lazy evaluation, high-order logic \\ \addlinespace[2pt]
\makecell[l]{\textbf{attrs} \\ \cite{attrs_repo}}     & 6.3k & 209 & Metaprogramming, class gen, attribute validation \\ \addlinespace[2pt]
\makecell[l]{\textbf{jinja2} \\ \cite{jinja2_repo}}   & 14k & 527 & Template compilation, lexical parsing, context flow \\ \addlinespace[2pt]
\makecell[l]{\textbf{networkx} \\ \cite{networkx_repo}} & 67k+ & 534 & Graph traversal, centrality, complex connectivity \\ \addlinespace[2pt]
\makecell[l]{\textbf{sympy} \\ \cite{sympy_repo}}     & 776k & 913 & Symbolic math, solver logic, recursive calculus \\
\bottomrule
\end{tabularx}
\end{table}

\section{Other Statistics from Experiment and Dataset}

\subsection{Diagnostic Insights: Reasoning Scope}

\begin{figure}
    \centering
    \includegraphics[width=\columnwidth]{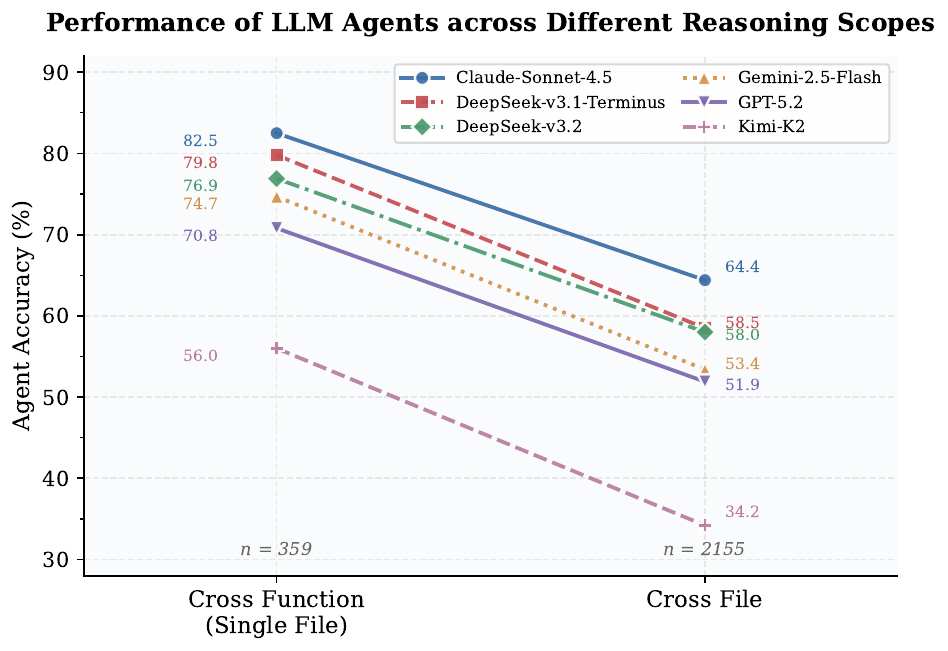}
    \caption{Performance of LLM agents across different reasoning scopes.}
    \label{fig:scope}
\end{figure}

Figure~\ref{fig:scope} illustrates the impact of reasoning scope on agent performance. We observe a consistent performance decay as the task transitions from Intra-file Isolation to Repository-Wide Synthesis. This downward trend quantifies the Retrieval Failure described in our diagnostic framework, highlighting a significant Spatial Deficit in current agents' ability to locate and navigate long-distance dependencies across the repository architecture. Notably, cross-file reasoning tasks constitute the vast majority of RepoReason, further emphasizing its authentic repository-level nature.

\subsection{Dataset Characteristics: Cognitive Metrics Distribution}

\begin{figure*}[htbp]
\centering
\includegraphics[width=\textwidth]{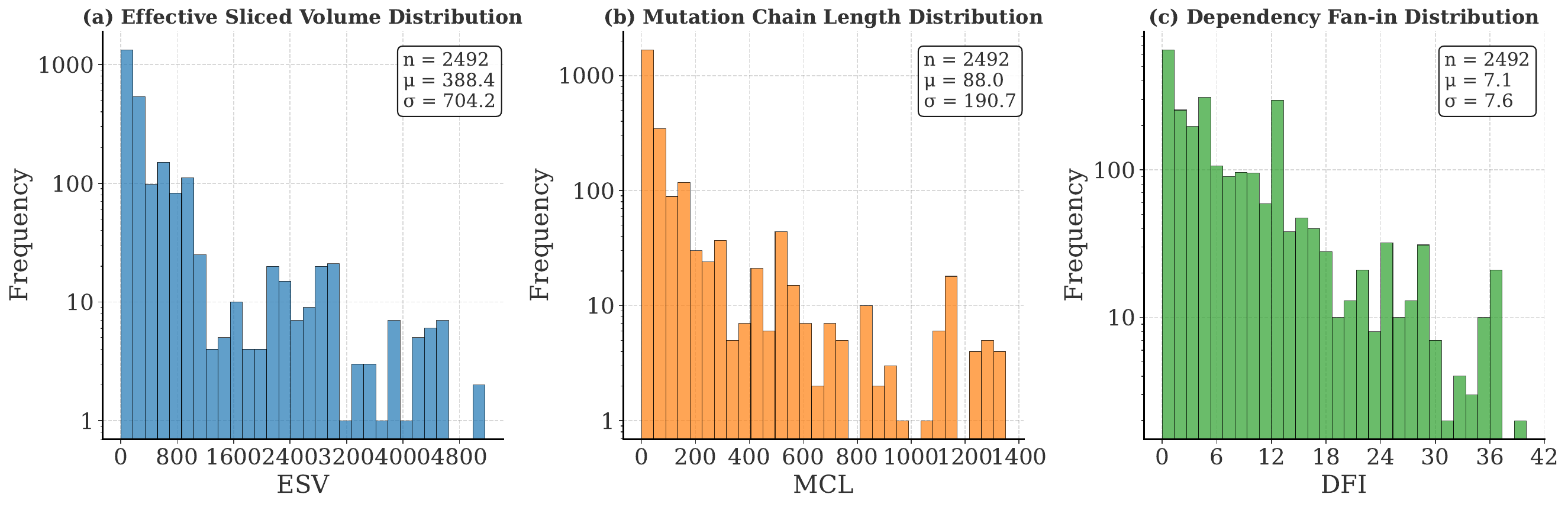}
\caption{Empirical distribution of the three cognitive metrics across the $\text{RepoReason}$ dataset: (a) $\text{ESV}$, (b) $\text{MCL}$, and (c) $\text{DFI}$.}
\label{fig:metrics_distribution}
\end{figure*}

Figure~\ref{fig:metrics_distribution} illustrates the empirical distribution of the three cognitive metrics across the $\text{RepoReason}$ dataset. All metrics exhibit a characteristic long-tail distribution, indicating a wide range of task difficulties. The Effective Sliced Volume ($\text{ESV}$) distribution ($\mu=393.6, \sigma=706.3$) shows that a significant portion of tasks require navigating hundreds or thousands of lines of code. The Mutation Chain Length ($\text{MCL}$) distribution ($\mu=93.9, \sigma=198.0$) highlights the depth of state tracking required, while the Dependency Fan-in ($\text{DFI}$) distribution ($\mu=8.3, \sigma=7.5$) captures the integration complexity of multi-source information. This diverse distribution ensures that $\text{RepoReason}$ provides sufficient data points across various complexity levels for fine-grained diagnostic analysis.

\subsection{Partial Correlation Analysis of Cognitive Metrics}

While ESV, MCL, and DFI exhibit conceptual orthogonality (measuring reading comprehension, sequential simulation, and multi-source synthesis, respectively), they naturally correlate in real-world codebases where complex tasks demand simultaneous scaling across all cognitive dimensions. To isolate the independent contribution of each metric, we conducted a partial correlation analysis on GPT-5.2 accuracy ($N=2492$), controlling for the other two metrics in each case (Table~\ref{tab:partial_corr}).

\begin{table}[htbp]
\footnotesize
\centering
\renewcommand{\arraystretch}{1.25}
\caption{Partial Correlation of Each Metric with Accuracy, Controlling for the Other Two ($N=2492$, GPT-5.2)}
\label{tab:partial_corr}
\setlength{\tabcolsep}{3pt}
\begin{tabularx}{\columnwidth}{@{} l l >{\centering\arraybackslash}X >{\centering\arraybackslash}X @{}}
\toprule
\textbf{Metric} & \textbf{Controlled For} & \textbf{Partial $r$} & \textbf{$p$-value} \\
\midrule
\rowcolor[HTML]{F5F5F5}
ESV & MCL, DFI & $-0.004$ & $0.835$ \\
MCL & ESV, DFI & $-0.021$ & $0.310$ \\
\rowcolor[HTML]{FFFDE7}
DFI & ESV, MCL & $\mathbf{-0.127}$ & $\mathbf{1.03 \times 10^{-9}}$ \\
\bottomrule
\end{tabularx}
\end{table}

After controlling for ESV and MCL, DFI maintains an independent and highly significant negative correlation with accuracy ($r = -0.127, p < 10^{-9}$), whereas the partial correlations for ESV ($r = -0.004$) and MCL ($r = -0.021$) become statistically insignificant. This confirms that integration width constitutes a unique cognitive bottleneck distinct from general task difficulty or reading load.

\subsection{Robustness Analysis: Temperature Sensitivity}

To verify the stability of our findings against generation hyperparameters, we conducted a supplementary experiment using Claude-Sonnet-4.5 under both greedy decoding (\texttt{temperature=0}) and default sampling (\texttt{temperature=1.0}). Table~\ref{tab:temp_acc} presents the repository-level accuracy changes. The overall accuracy shift is minimal ($+1.00$pp). Crucially, the Pearson correlations between accuracy and the three cognitive metrics remain remarkably stable (Table~\ref{tab:temp_corr}), demonstrating that the observed aggregation deficit and cliff effects are structural failures of the model architecture, rather than artifacts of decoding strategies.

\begin{table}[htbp]
\footnotesize
\centering
\renewcommand{\arraystretch}{1.25}
\caption{Claude-Sonnet-4.5 Accuracy (\%): \texttt{temperature=0} vs.\ \texttt{1.0}}
\label{tab:temp_acc}
\setlength{\tabcolsep}{4pt}
\begin{tabularx}{\columnwidth}{@{} l *{3}{>{\centering\arraybackslash}X} @{}}
\toprule
\textbf{Repository} & \textbf{$T=0$} & \textbf{$T=1.0$} & \textbf{$\Delta$ (pp)} \\
\midrule
\rowcolor[HTML]{F5F5F5}
\texttt{attrs}      & 72.41 & 74.14 & $+1.73$ \\
\texttt{cachetools} & 79.79 & 79.79 & $0.00$ \\
\rowcolor[HTML]{F5F5F5}
\texttt{jinja2}     & 67.46 & 68.92 & $+1.46$ \\
\texttt{networkx}   & 79.77 & 80.20 & $+0.43$ \\
\rowcolor[HTML]{F5F5F5}
\texttt{sympy}      & 51.10 & 52.70 & $+1.60$ \\
\texttt{toolz}      & 82.81 & 83.44 & $+0.63$ \\
\rowcolor[HTML]{F5F5F5}
\texttt{yarl}       & 78.65 & 80.73 & $+2.08$ \\
\midrule
\rowcolor[HTML]{E8F5E9}
\textbf{Overall}    & \textbf{66.98} & \textbf{67.98} & \textbf{$+1.00$} \\
\bottomrule
\end{tabularx}
\end{table}

\FloatBarrier

\begin{table}[htbp]
\footnotesize
\centering
\renewcommand{\arraystretch}{1.25}
\caption{Stability of Metric-Accuracy Pearson $r$}
\label{tab:temp_corr}
\setlength{\tabcolsep}{4pt}
\begin{tabularx}{\columnwidth}{@{} l *{3}{>{\centering\arraybackslash}X} @{}}
\toprule
\textbf{Correlation} & \textbf{$T=0$} & \textbf{$T=1.0$} & \textbf{$\Delta$} \\
\midrule
\rowcolor[HTML]{F5F5F5}
ESV -- Accuracy & $-0.158$ & $-0.170$ & $-0.012$ \\
MCL -- Accuracy & $-0.137$ & $-0.142$ & $-0.005$ \\
\rowcolor[HTML]{F5F5F5}
DFI -- Accuracy & $-0.218$ & $-0.229$ & $-0.011$ \\
\bottomrule
\end{tabularx}
\end{table}

\section{Technical Specification of the Agentic Search Framework}

This section details the implementation of the Agentic Search evaluation framework, focusing on the specialized prompt engineering for the \texttt{ReadOnlyAgent} and the robust evaluation mechanisms used to verify repository-level reasoning.

\subsection{Agentic Search Prompting Strategy}
The framework utilizes a template-driven approach to construct tasks for the Agent. As illustrated in Box 1, the prompt is structured to provide necessary execution context while enforcing strict output constraints.

\begin{agentbox}[Agentic Search Task Prompt Template]
\small
\textbf{Task:} Fill in the blank in the Python test code. \\

\textbf{Blank Assertion:} \\
\texttt{[Blanked\_Assertion\_Statement]} \\

\textbf{Hint:} \\
\textit{[Type\_Specific\_Hint]} (e.g., "The answer MUST be a \textbf{string literal}" or "MUST be a \textbf{constructor call} from the source code"). \\

\textbf{Instructions:} \\
The '\_\_\_' in the assertion represents a missing expression (variable name, function call, or literal). \\

You are a \texttt{ReadOnlyAgent}. You can read files and search the code in the repository to infer what this expression should be. \\
Note: Test function code is not visible in the repository, but you can explore the source code functions called by the test and reason to complete the fill-in-the-blank task. \\

\textbf{Context:} \\
- \textbf{Test function source code:}
\begin{lstlisting}[language=Python]
[Source_Code_with_Question_Marker]
\end{lstlisting}
- \textbf{Input parameters:} \texttt{[JSON\_Serialized\_Inputs]}

Note: The missing expression is likely related to the input parameters or variables defined in the test or the code being tested. \\

\textbf{Goal:} \\
Find the exact expression that should replace '\_\_\_'. \\
When you are confident, use the \texttt{finish} tool with the answer. \\

\textbf{IMPORTANT:} The finish tool's content MUST contain \textbf{ONLY} the answer expression itself, nothing else. Examples:
\begin{itemize}[leftmargin=*,noitemsep,topsep=2pt]
    \item If the answer is a variable: finish with content \texttt{"result"}
    \item If the answer is a function call: finish with content \texttt{"len(items)"}
    \item If the answer is a string: finish with content \texttt{'"hello"'}
    \item If the answer is an attribute access: finish with content \texttt{"C.\_\_attrs\_\_attrs\_\_[1].validator"}
\end{itemize}

Do NOT include explanations, reasoning, or any other text in the finish tool content. Only the answer expression.
\end{agentbox}

\subsubsection{Concrete Example: Literal Integer Answer}
\label{app:prompt-example}

We present a real prompt instance from the cachetools repository to illustrate the complete Agentic Search prompt structure. This example demonstrates a literal integer answer (not a variable reference), where the Agent must infer the value \texttt{2} by reasoning about cache operations.

\begin{agentbox}[Real Agentic Search Prompt Example]
\small
\textbf{Task:} Fill in the blank in the Python test code. \\

\textbf{Blank Assertion:} \\
\texttt{assert \_\_\_ == len(cache)} \\

\textbf{Hint:} \\
\textbf{CRITICAL REQUIREMENT}: The answer MUST be a literal integer value. \\

IMPORTANT: Return ONLY the literal value, NOT an expression or variable name. \\

Type-specific examples:
\begin{itemize}[leftmargin=*,noitemsep,topsep=2pt]
    \item Boolean: True or False (NOT 'x == y' or 'bool(x)')
    \item String: "hello" (NOT 'str\_var' or 'x.name')
    \item Number: 42 (NOT 'num\_var' or 'len(x)')
    \item List: [1, 2, 3] (NOT 'list\_var' or 'list(x)')
    \item Dict: \{"a": 1\} (NOT 'dict\_var' or 'dict(x)')
\end{itemize}

\textbf{Instructions:} \\
The '\_\_\_' in the assertion represents a missing expression (variable name, function call, or literal). \\

You are a \texttt{ReadOnlyAgent}. You can read files and search the code in the repository to infer what this expression should be. \\
Note: Test function code is not visible in the repository, but you can explore the source code functions called by the test and reason to complete the fill-in-the-blank task. \\

\textbf{Context:} \\
- \textbf{Test function source code:}
\begin{lstlisting}[language=Python]
def test_update(self):
    cache = self.Cache(maxsize=2)

    cache.update({1: 1, 2: 2})
    assert ___ == len(cache)  # Question
    assert ___ == cache[1]
    assert ___ == cache[2]

    cache.update({1: 1, 2: 2})
    assert ___ == len(cache)
    assert ___ == cache[1]
    assert ___ == cache[2]

    cache.update({1: "a", 2: "b"})
    assert ___ == len(cache)
    assert ___ == cache[1]
    assert ___ == cache[2]
\end{lstlisting}
- \textbf{Input parameters:} \texttt{\{"self": "<LRUCacheTest>"\}} \\

Note: The missing expression is likely related to the input parameters or variables defined in the test or the code being tested. \\

\textbf{Goal:} \\
Find the exact expression that should replace '\_\_\_'. \\
When you are confident, use the \texttt{finish} tool with the answer. \\

\textbf{IMPORTANT:} The finish tool's content MUST contain \textbf{ONLY} the answer expression itself, nothing else. Examples:
\begin{itemize}[leftmargin=*,noitemsep,topsep=2pt]
    \item If the answer is a variable: finish with content \texttt{"result"}
    \item If the answer is a function call: finish with content \texttt{"len(items)"}
    \item If the answer is a string: finish with content \texttt{'"hello"'}
    \item If the answer is an attribute access: finish with content \texttt{"C.\_\_attrs\_\_attrs\_\_[1].validator"}
\end{itemize}

Do NOT include explanations, reasoning, or any other text in the finish tool content. Only the answer expression.
\end{agentbox}

\noindent\textbf{Analysis.}
In this example, the Agent must:
\begin{enumerate}[leftmargin=*,noitemsep,topsep=2pt]
    \item Understand that \texttt{cache.update(\{1: 1, 2: 2\})} adds two key-value pairs to a cache with \texttt{maxsize=2}
    \item Reason that after the update, \texttt{len(cache)} equals \texttt{2}
    \item Infer that the blank should be filled with the literal integer \texttt{2}
    \item Use the \texttt{finish} tool with content \texttt{"2"} (without quotes in the tool content, as it's a numeric literal)
\end{enumerate}

This example demonstrates how the prompt guides the Agent to reason about runtime behavior while enforcing strict output constraints for evaluation.

\subsection{Evaluation and Comparison Engine}
Given the syntactic flexibility of Python, the framework employs a multi-criteria comparison engine to verify the Agent's output against ground truth.

\subsubsection{Mathematical Equivalence (\texorpdfstring{$\mathcal{M}$}{M})}
For tasks involving symbolic expressions, the engine utilizes the \texttt{SymPy} library to verify mathematical identity. Correctness is defined as:
\begin{equation}
\begin{aligned}
\text{Correct} \iff & \text{Simplify}(P - G) = 0 \\
& \lor \text{Expand}(P) = \text{Expand}(G)
\end{aligned}
\end{equation}
where $P$ is the Agent's prediction and $G$ is the ground truth. This multi-path verification prevents false negatives caused by structural variations.

\subsubsection{Structural Normalization (\texorpdfstring{$\mathcal{S}$}{S})}
The engine applies recursive normalization to handle implementation-specific data structures:
\begin{itemize}[leftmargin=*,noitemsep,topsep=2pt]
    \item \textbf{Wrapper Stripping:} Automatically unwrap custom container types (e.g., \texttt{D(\{...\})} $\to$ \texttt{\{...\}}).
    \item \textbf{Collection Alignment:} Standardize whitespace and element ordering in complex literals.
    \item \textbf{Object Mapping:} Map specialized collections like \texttt{defaultdict} to standard dictionary equivalents.
\end{itemize}

\subsubsection{Heuristic Semantic Matching}
To bridge the gap between precise syntax and reasoning intent, the following heuristics are applied:
\begin{itemize}[leftmargin=*,noitemsep,topsep=2pt]
    \item \textbf{Constructor Awareness:} Recognizes class instantiations required by the logic (e.g., \texttt{ClassName(args)}).
    \item \textbf{Path Qualification:} Standardizes module paths, accepting partially qualified references if unambiguous.
    \item \textbf{Quote Invariance:} Neutralizes differences in quote types (single, double, or triple).
\end{itemize}

\subsection{Execution Management}
The Agent is executed in a sandboxed environment with a strict 1800-second timeout. All tool invocations are captured in a serialized JSON sequence for auditability. An incremental checkpointing mechanism is implemented to ensure data persistence during large-scale evaluations.

\section{Countering Memorization via an Execution-Driven Mutation Engine}
\label{app:edme}

This appendix provides detailed implementation specifications to ensure full reproducibility of our Execution-Driven Mutation Engine (EDME). We present the system's I/O contracts, hyperparameters, algorithmic components, and concrete examples from real mutation instances.

\subsection{Reproducibility Artifacts and I/O Contracts}
\label{app:edme-io}

\noindent\textbf{Artifacts.}
Each mutation instance is serialized as a JSON record containing: (i) the original test excerpt, (ii) the probe-injected intermediate code, (iii) captured runtime values, (iv) the final mutated test, and (v) validation metadata (executability, assertion validity, and API preservation).

\noindent\textbf{Input Contract.}
The system takes as input:
\begin{itemize}[leftmargin=*,noitemsep,topsep=2pt]
\item A unit-test function (source text)
\item Extracted API call signature list (Section~\ref{app:edme-api})
\item Original assertion list with their structural patterns
\item Repository-specific execution environment configuration
\end{itemize}

\noindent\textbf{Output Contract.}
The system emits a mutated test function that satisfies three mandatory properties:
\begin{enumerate}[leftmargin=*,noitemsep,topsep=2pt]
\item \mbox{\textbf{Executability}}: runs without errors in the target environment
\item \mbox{\textbf{Assertion Validity}}: all reconstructed assertions pass
\item \mbox{\textbf{API Preservation}}: maintains the original API call sequence
\end{enumerate}

\noindent\textbf{Metadata Schema.}
Each output record includes:
\begin{itemize}[leftmargin=*,noitemsep,topsep=2pt]
\item \mbox{\texttt{validation\_status}}: \texttt{valid} or \texttt{invalid}
\item \mbox{\texttt{api\_calls\_preserved}}: boolean flag
\item \mbox{\texttt{execution\_driven\_assert}}: nested object with probe results
\item \mbox{\texttt{mutation\_attempts}}: number of LLM retries
\item \mbox{\texttt{execution\_time}}: total processing time in seconds
\end{itemize}

\noindent\textbf{Output Structure.}
\begin{itemize}[leftmargin=*,noitemsep,topsep=2pt]
\item Structured artifacts capturing per-repository mutation records (JSON)
\item Aggregated statistics summarizing successful mutations
\item Logs containing execution traces and debug output
\end{itemize}

\subsection{Detailed Hyperparameters and Constraints}
\label{app:edme-hparams}

Table~\ref{tab:edme-hparams} lists the key hyperparameters and hard constraints required for faithful reproduction.

\begin{table}[t]
\scriptsize
\centering
\caption{Key hyperparameters and hard constraints for EDME.}
\label{tab:edme-hparams}
\setlength{\tabcolsep}{3pt}
\begin{tabular}{p{0.30\columnwidth} p{0.64\columnwidth}}
\toprule
\textbf{Item} & \textbf{Setting / Constraint} \\
\midrule
Teacher model & \texttt{gpt-5.2} (single-step unified mutation) \\
Temperature & 0.7 (balancing creativity and correctness) \\
Max tokens & 16384 per mutation request \\
Probe format & \textit{print(f"DEBUG\_RESULT: \{expr\}")} (mandatory\allowbreak{} f-string) \\
Max mutation retries & 3 (error-feedback loop) \\
Execution timeout & 180s per trial \\
Validation gate & All of: executability, assertion validity, API preservation \\
\midrule
\multicolumn{2}{l}{\textit{Hard Constraints (enforced via prompt):}} \\
No signature drift & Function signature must not change (no new args/defaults) \\
No new imports & Must reuse existing imports/helpers in the file \\
Preserve parametrize & \textit{pytest.mark.parametrize} decorators must remain\allowbreak{} intact \\
Style consistency & Match original assertion patterns (e.g., \texttt{is} \textit{vs.} \texttt{==}) \\
\bottomrule
\end{tabular}
\end{table}

\subsection{API-Call Sequence Preservation Mechanism}
\label{app:edme-api}

To prevent shortcutting and preserve the reasoning pathway, we enforce that the mutated test retains the original API call sequence to the target library.

\noindent\textbf{Extraction Phase.}
We traverse the original test's Abstract Syntax Tree (AST) to extract a list of call signatures:
\begin{itemize}[leftmargin=*,noitemsep,topsep=2pt]
\item Free function calls: \mbox{\texttt{solve(...)}}, \mbox{\texttt{simplify(...)}}
\item Method calls: \mbox{\texttt{obj.method(...)}}, \mbox{\texttt{expr.rewrite(...)}}
\item Constructor calls: \mbox{\texttt{FiniteSet(...)}}, \mbox{\texttt{Rational(...)}}
\end{itemize}
This extraction is performed by traversing the AST and collecting all callable expressions.

\noindent\textbf{Validation Phase.}
During validation, we check that each recorded call signature appears in the mutated code as a callable occurrence. The system constructs a regex pattern for each call (matching the function/method name followed by an opening parenthesis) and verifies its presence. Any candidate that removes, substitutes, or reorders these calls is rejected.

\noindent\textbf{Result Recording.}
The validation result is written to the metadata. Only mutations that preserve API calls proceed to the final output dataset.

\subsection{Execution-Driven Mutation and Assertion Reconstruction}
\label{app:edme-assert}

EDME employs a three-stage pipeline that first generates mutated code with embedded probes, then executes it to capture ground-truth values, and finally reconstructs assertions from the captured runtime values. This approach eliminates model hallucination by deriving truth from actual code execution.

\subsubsection{Stage 1: Comprehensive Mutation with Probe Generation}
\label{app:edme-assert-probe}

In a single LLM call, the system performs both visual/logical mutations and probe injection simultaneously. The LLM is instructed to generate mutated code that directly replaces all original assertions with \texttt{DEBUG\_RESULT} print statements, combining mutation and probe generation in one unified step.

\noindent\textbf{Comprehensive Mutation Process.}
The mutation prompt requires the LLM to perform two tasks in parallel:
\begin{enumerate}[leftmargin=*,noitemsep,topsep=2pt]
\item \mbox{\textbf{Visual mutations}}: Rename local variables, restructure control flow, change visual appearance
\item \mbox{\textbf{Logical mutations}}: Change literal constants, modify input parameters, perturb data flow
\item \mbox{\textbf{Probe injection}}: Replace each assertion of the form \textit{assert expression == expected\_value} with \textit{print(f"DEBUG\_RESULT: \{expression\}")}
\end{enumerate}
This unified approach ensures that the mutated code is immediately executable as a state probe, with all mutations and probe placements completed in a single LLM generation step.

\noindent\textbf{Format Enforcement.}
The probe format is strictly enforced via prompt constraints and post-generation validation:
\begin{itemize}[leftmargin=*,noitemsep,topsep=2pt]
\item Must use f-string syntax (not \textit{print("DEBUG\_RESULT:", x)})
\item Must include the exact prefix \mbox{\texttt{DEBUG\_RESULT:}}
\item Multi-line expressions must be properly formatted
\end{itemize}

\noindent\textbf{Automatic Correction.}
If the LLM generates incorrect probe format (e.g., using comma-separated print), the system automatically corrects it via regex replacement. If the generated code lacks \texttt{DEBUG\_RESULT} outputs, the system triggers a retry with explicit feedback.

\subsubsection{Stage 2: Execution-Driven Value Capture}
\label{app:edme-assert-runtime}

After generating the mutated code with embedded probes, we execute it in a controlled environment and parse the stdout stream to extract \texttt{DEBUG\_RESULT} payloads. This execution step captures the actual runtime values produced by the mutated logic, which serve as ground truth for assertion reconstruction.

\noindent\textbf{Execution Environment.}
The execution is performed with the following setup:
\begin{itemize}[leftmargin=*,noitemsep,topsep=2pt]
\item Isolated temporary file in the original test directory (to support relative imports)
\item Repository-specific virtual environment (configured for the target project)
\item Configured \mbox{\texttt{PYTHONPATH}} to include necessary modules
\item 180-second timeout per execution
\end{itemize}

\noindent\textbf{Output Parsing.}
The parser handles multiple output formats:
\begin{enumerate}[leftmargin=*,noitemsep,topsep=2pt]
\item Standalone line: \texttt{DEBUG\_RESULT: value}
\item Embedded in pytest output: \textit{... DEBUG\_RESULT: value ...}
\item Multi-line structures: continues collecting lines until brackets/braces are balanced
\end{enumerate}

\noindent\textbf{Strict Mode.}
During probe execution, \textit{check\_debug\_result=True} enforces that at least one \texttt{DEBUG\_RESULT} must be found. Execution success without probe output is treated as failure, triggering a retry with error feedback.

\subsubsection{Stage 3: Assertion Reconstruction from Runtime Values}
\label{app:edme-assert-style}

Using the captured runtime values from Stage 2 as ground truth, a second LLM call regenerates assertions while matching the original test's idioms. The LLM converts each \texttt{DEBUG\_RESULT} output back into an appropriate assertion statement, ensuring style consistency with the original test.

\noindent\textbf{Style Matching Rules.}
The LLM is instructed to preserve:
\begin{itemize}[leftmargin=*,noitemsep,topsep=2pt]
\item Comparison operators: \mbox{\texttt{is}} \textit{vs.} \mbox{\texttt{==}} \textit{vs.} \mbox{\texttt{!=}}
\item Constructor preferences: \mbox{\texttt{S.Half}} \textit{vs.} \mbox{\texttt{0.5}}, \mbox{\texttt{Rational(a,b)}} \textit{vs.} \mbox{\texttt{a/b}}
\item Collection types: \mbox{\texttt{FiniteSet(...)}} \textit{vs.} \mbox{\texttt{\{...\}}}, \mbox{\texttt{[...]}} \textit{vs.} \mbox{\texttt{(...)}}
\item Assertion structure: single-line \textit{vs.} multi-line, with/without intermediate variables
\end{itemize}

\noindent\textbf{Error-Driven Refinement.}
The assertion reconstruction supports iterative refinement:
\begin{enumerate}[leftmargin=*,noitemsep,topsep=2pt]
\item Generate assertions from captured runtime values
\item Replace \mbox{\texttt{DEBUG\_RESULT}} statements with the generated assertions
\item Execute the final mutated code to validate all assertions pass
\item If validation fails, append error message to conversation history
\item Retry assertion generation with error feedback (up to 3 attempts)
\end{enumerate}

\subsection{Strict Validation Gate}
\label{app:edme-gate}

A mutated test is accepted \emph{only if} it satisfies all three validation criteria simultaneously.

\noindent\textbf{Criterion 1: Executability.}
The probe-injected code must:
\begin{itemize}[leftmargin=*,noitemsep,topsep=2pt]
\item Run without syntax errors or runtime exceptions
\item Complete within the 180-second timeout
\item Produce at least one \mbox{\texttt{DEBUG\_RESULT}} output
\end{itemize}

\noindent\textbf{Criterion 2: Assertion Validity.}
The final mutated code with reconstructed assertions must:
\begin{itemize}[leftmargin=*,noitemsep,topsep=2pt]
\item Execute successfully (pytest returncode = 0)
\item Pass all assertions against the mutated logic
\end{itemize}

\noindent\textbf{Criterion 3: API Preservation.}
The mutated code must:
\begin{itemize}[leftmargin=*,noitemsep,topsep=2pt]
\item Contain all original API call signatures
\item Maintain the same call sequence (order may vary within independent statements)
\end{itemize}

\noindent\textbf{Gate Enforcement.}
The validation status is recorded in metadata as \texttt{validation\_status}. Only mutations with \texttt{validation\_status="valid"} are included in the final dataset. The batch validation routine skips non-valid entries during final verification.

\subsection{Key Prompt Templates}
\label{app:edme-prompts}

This subsection presents the essential prompt templates used in EDME's LLM-based mutation and assertion reconstruction stages.

\subsubsection{Mutation Prompt (Stage 1)}
\label{app:edme-prompt-mutation}

The mutation prompt enforces hard constraints while guiding the model to perform both visual and logical mutations. The system prompt emphasizes:

\noindent\textbf{System Prompt (Excerpt).}
\begin{itemize}[leftmargin=*,noitemsep,topsep=2pt]
\item \mbox{\textbf{Format requirements}}: All \texttt{DEBUG\_RESULT} outputs must use f-string format: \textit{print(f"DEBUG\_RESULT: \{value\}")}
\item \mbox{\textbf{Visual mutations}}: Rename local variables, restructure control flow, change visual appearance
\item \mbox{\textbf{Logical mutations}}: Change literal constants, modify input parameters, perturb data flow
\item \mbox{\textbf{API preservation}}: Maintain the exact API call sequence specified in constraints
\item \mbox{\textbf{Signature preservation}}: Function signature must remain unchanged (no new args/defaults)
\item \mbox{\textbf{Import reuse}}: Do not add new imports; reuse existing imports/helpers
\end{itemize}

\noindent\textbf{User Prompt (Excerpt).}
The user prompt provides:
\begin{itemize}[leftmargin=*,noitemsep,topsep=2pt]
\item Target function name and context (decorators, class definition if applicable)
\item API call constraints list (must be strictly preserved)
\item Original test code with all assertions
\item Available imports and helper functions
\item Explicit instruction to replace all assertions with \mbox{\texttt{DEBUG\_RESULT}} outputs
\end{itemize}

\subsubsection{Assertion Reconstruction Prompt (Stage 3)}
\label{app:edme-prompt-reconstruct}

The assertion reconstruction prompt guides the model to convert captured runtime values into style-consistent assertions.

\noindent\textbf{System Prompt (Excerpt).}
\begin{itemize}[leftmargin=*,noitemsep,topsep=2pt]
\item \mbox{\textbf{Style matching}}: Strictly follow original assertion patterns (operators, constructors, constants)
\item \mbox{\textbf{Format preservation}}: Maintain \texttt{is} \textit{vs.} \texttt{==}, \texttt{Rational(a,b)} \textit{vs.} \texttt{a/b}, collection types
\item \mbox{\textbf{No markdown}}: Return pure Python code only, no explanations
\item \mbox{\textbf{Type conversion}}: Parse runtime values and convert to original style (e.g., \texttt{Rational(2, 3)} instead of \texttt{0.666...})
\end{itemize}

\noindent\textbf{User Prompt (Excerpt).}
The user prompt provides:
\begin{itemize}[leftmargin=*,noitemsep,topsep=2pt]
\item Captured runtime values (from \mbox{\texttt{DEBUG\_RESULT}} outputs) as JSON array
\item Original assertion patterns (for style reference)
\item Code with \mbox{\texttt{DEBUG\_RESULT}} statements to convert
\item Available imports context
\item Error feedback (if previous attempt failed)
\end{itemize}

\subsection{End-to-End Algorithm}
\label{app:edme-pseudo}

Algorithm~\ref{alg:edme} presents the complete EDME pipeline in pseudocode.

\begin{algorithm}[t]
\caption{Execution-Driven Mutation Engine (EDME)}
\label{alg:edme}
\footnotesize
\begin{algorithmic}[1]
\Require Original test $T$, API inventory $\mathcal{A}$,
\Statex \hspace{2em}configuration $C$
\Ensure Mutated test $T'$ with metadata $M$, or $\bot$
\Statex \hspace{2em}if no candidate passes validation
\State $\textit{attempts} \gets 0$
\While{$\textit{attempts} < C.\textit{max\_retries}$}
  \State $\textit{attempts} \gets \textit{attempts} + 1$
  \algstage{Stage 1: Probe-aware mutation}
  \State $T_{\text{probe}} \gets \textsc{MutateWithProbes}(T, \mathcal{A})$
  \If{$\neg\, \textsc{HasDebugProbe}(T_{\text{probe}})$}
    \State \textbf{continue} \Comment{Missing probe}
  \EndIf
  \algstage{Stage 2: Runtime capture}
  \State $(\textit{exec}, V) \gets \textsc{ExecuteProbes}(T_{\text{probe}},$
  \Statex \hspace{2em}$C.\textit{timeout})$
  \If{$\neg\, \textit{exec}$}
    \State \textbf{continue} \Comment{Execution failed}
  \EndIf
  \algstage{Stage 3: Assertion reconstruction}
  \State $T' \gets \textsc{ReconstructAssertions}(T_{\text{probe}}, V,$
  \Statex \hspace{2em}$\textsc{StyleOf}(T))$
  \State $\textit{assert\_ok} \gets \textsc{ExecuteFinal}(T',$
  \Statex \hspace{2em}$C.\textit{timeout})$
  \If{$\neg\, \textit{assert\_ok}$}
    \State \textbf{continue} \Comment{Assertion validation failed}
  \EndIf
  \State $\textit{api\_ok} \gets \textsc{VerifyAPI}(T, T', \mathcal{A})$
  \If{$\neg\, \textit{api\_ok}$}
    \State \textbf{continue} \Comment{API preservation failed}
  \EndIf
  \State $M \gets \textsc{RecordMetadata}(\texttt{valid},$
\Statex \hspace{2em}$\textit{api\_ok}, V)$
  \State \Return $(T', M)$
\EndWhile
\State $M \gets \textsc{RecordMetadata}(\texttt{invalid},$
\Statex \hspace{2em}$\textbf{false}, \emptyset)$
\State \Return $(\bot, M)$
\end{algorithmic}
\end{algorithm}

\subsection{Concrete Example from Real Data}
\label{app:edme-example}

We present a real mutation instance from our dataset to illustrate the complete EDME pipeline. This example is taken from the SymPy repository, specifically the \texttt{test\_rewrite\_trig} function from the solvers test suite.

This instance demonstrates:
\begin{itemize}[leftmargin=*,noitemsep,topsep=2pt]
\item Fact-level perturbation (changing coefficients from $2, 1$ to $3, 2$)
\item API preservation (\mbox{\texttt{solve}}, \mbox{\texttt{sin}}, \mbox{\texttt{cos}}, \mbox{\texttt{atan}} all retained)
\item Execution-driven truth (runtime value \mbox{\texttt{[atan(2/3)]}} captured)
\item Style consistency (using \mbox{\texttt{Rational(2, 3)}} instead of \mbox{\texttt{0.666...}})
\end{itemize}

\begin{figure*}[t]
\centering
\caption{Real EDME instance from SymPy. Left: original test with memorized constants. Middle: probe-injected candidate with mutated coefficients and structured debug output. Right: final assertion reconstructed from runtime ground truth while keeping the preserved API pathway.}
\label{fig:edme-example}
\begin{subfigure}[t]{0.3\textwidth}
\caption*{Original Test}
\begin{lstlisting}[style=tracefig]
def test_rewrite_trig():
    # Original equation
    assert solve(
        2*sin(x) - cos(x),
        x
    ) == [atan(S.Half)]
\end{lstlisting}
\end{subfigure}\hfill
\begin{subfigure}[t]{0.32\textwidth}
\caption*{Probe-Injected}
\begin{lstlisting}[style=tracefig]
def test_rewrite_trig():
    # Mutated equation
    probe = x*(S.One + S.Zero)
    print(
        f"DEBUG_RESULT: {solve(3*sin(probe) - 2*cos(probe), probe)}"
    )
\end{lstlisting}
\end{subfigure}\hfill
\begin{subfigure}[t]{0.3\textwidth}
\caption*{Final Mutated}
\begin{lstlisting}[style=tracefig]
def test_rewrite_trig():
    # Reconstructed assertion
    probe = x*(S.One + S.Zero)
    assert solve(
        3*sin(probe) - 2*cos(probe),
        probe
    ) == [atan(Rational(2, 3))]
\end{lstlisting}
\end{subfigure}
\end{figure*}

\noindent\textbf{Mutation Analysis.}
The key transformations in this example:
\begin{enumerate}[leftmargin=*,noitemsep,topsep=2pt]
\item \mbox{\textbf{Visual mutation}}: Variable $x$ renamed to $\text{probe} = x \cdot (S.\text{One} + S.\text{Zero})$ (semantically equivalent but visually distinct)
\item \mbox{\textbf{Logic mutation}}: Coefficients changed from $(2, -1)$ to $(3, -2)$, invalidating the original answer $\arctan(1/2)$
\item \mbox{\textbf{Probe execution}}: Captured runtime output \texttt{[atan(2/3)]} from the mutated equation
\item \mbox{\textbf{Style preservation}}: Reconstructed as \texttt{Rational(2, 3)} (matching SymPy's preference for exact rationals) instead of \texttt{0.6666...}
\end{enumerate}

\noindent\textbf{Validation Results.}
This mutation passed all three validation criteria:
\begin{itemize}[leftmargin=*,noitemsep,topsep=2pt]
\item \mbox{\textbf{Executability}}: Probe code ran successfully, producing \textit{DEBUG\_RESULT: [atan(2/3)]}
\item \mbox{\textbf{Assertion validity}}: Final code passed pytest with the reconstructed assertion
\item \mbox{\textbf{API preservation}}: All four API calls (\mbox{\texttt{solve}}, \mbox{\texttt{sin}}, \mbox{\texttt{cos}}, \mbox{\texttt{atan}}) present in original order
\end{itemize}

\lstset{style=dvp}

\section{Countering Ambiguity via the Deterministic Value Protocol}
\label{app:dvp}

This appendix describes the Deterministic Value Protocol (DVP). DVP enforces hard constraints on ground-truth answers during task generation. The goal is to remove evaluation noise caused by (i) unstable string representations, (ii) environment-dependent nondeterminism, and (iii) multi-solution semantics under strict matching.

\subsection{Artifacts and I/O Contracts}
\label{app:dvp-io}

\appsec{Input.}
DVP consumes one trace-derived test function and its assertions.

\appsec{Output.}
For each accepted assertion, DVP emits one fill-in-the-blank instance with the masked assertion and the ground-truth answer.

\appsec{Discard policy.}
If no mask position can satisfy DVP constraints, the sample is discarded.

\subsection{System Overview}
\label{app:dvp-overview}

DVP is programmatic-first and fallback-last. It applies two strict ``programmatic parsing\allowbreak{} funnels'' and uses an LLM fallback only under a narrow boundary.

\begin{itemize}[leftmargin=*,noitemsep,topsep=2pt]
\item \mbox{\textbf{Funnel 1 (Semantic Determinism):}} remove nondeterminism and multi-solution semantics.
\item \mbox{\textbf{Funnel 2 (Answer Morphological Determinism):}} keep only answer forms that are stable and explicit.
\end{itemize}

LLM fallback is invoked only when the assertion is structurally maskable but the extracted answer is a variable reference. In this case, the model is restricted to a \emph{variable resolver}. The resolved result must pass Funnel~1 and Funnel~2 again.

\subsection{Funnel 1: Semantic Determinism}
\label{app:dvp-f1}

Funnel~1 rejects candidates that can change across runs or admit multiple valid solutions.

\appsec{(1) Representational drift.}
DVP rejects unstable string forms. Typical patterns include:
\begin{itemize}[leftmargin=*,noitemsep,topsep=2pt]
\item object repr addresses: \mbox{\texttt{<... at 0x[0-9a-fA-F]+>}}
\item explicit memory addresses: \mbox{\texttt{0x[0-9a-fA-F]+}}
\end{itemize}

\appsec{(2) External perturbation.}
DVP rejects contexts that rely on runtime-dependent signals (randomness, timestamps, UUIDs). The detector checks tokens such as \texttt{random}, \texttt{uuid}, \texttt{time.time}, \texttt{datetime.now}, and \texttt{date.today}. If matched, the sample is discarded with a reason like \textit{nondeterministic\_in\_test:...}.

\appsec{(3) Multi-solution semantics.}
DVP rejects mask positions that admit multiple satisfying values under strict comparison. In operator masking, the system rejects \texttt{>}, \texttt{<}, \texttt{>=}, \texttt{<=}, and \texttt{!=}. For \texttt{==}, DVP also rejects reflexive variable forms (e.g., \texttt{x == y}).

\appsec{Approximate comparisons.}
DVP drops approximate floating comparisons that are sensitive to platform differences, including \texttt{pytest.approx}, \texttt{assertAlmostEqual}, \texttt{assert\_allclose}, \texttt{allclose}, and \texttt{isclose}.

\subsection{Funnel 2: Answer Morphological Determinism}
\label{app:dvp-f2}

Funnel~2 ensures the blank evaluates a concrete runtime state, not a name binding. The system assigns priorities to different answer forms (1 is best) or rejects them (priority 0), using AST-based type inference.

\appsec{Whitelist with priorities.}
Accepted answer forms are:
\begin{itemize}[leftmargin=*,noitemsep,topsep=2pt]
\item \mbox{\textbf{Priority 1: Literals}} (numeric, string, boolean constants, and container literals). Always accepted.
\item \mbox{\textbf{Priority 2: Global constants}} (uppercase identifiers following naming conventions).
\item \mbox{\textbf{Priority 3: Stable attribute access}} (e.g., \mbox{\texttt{response.\allowbreak status\_code}}, \mbox{\texttt{S.NaN}}), without call parentheses.
\item \mbox{\textbf{Priority 4: Parameter-resolvable constructors}} (\texttt{ClassName(...)}), only when all arguments are literals. Empty built-in constructors \texttt{set()}, \texttt{list()}, \texttt{dict()}, \texttt{tuple()}, \texttt{frozenset()} are treated as Priority~1.
\item \mbox{\textbf{Priority 5: Complex value expressions}} only for symbolic libraries (e.g., SymPy). As a special case, symbolic computation libraries like SymPy allow complex mathematical expressions (e.g., \textit{[Rational(33, 2) - Rational(11, 2)*sqrt(3)]}) as valid answers.
\end{itemize}

\appsec{Rejections.}
DVP rejects:
\begin{itemize}[leftmargin=*,noitemsep,topsep=2pt]
\item pure variable references
\item dynamic function calls (\mbox{\texttt{func()}} or \mbox{\texttt{obj.method()}}) unless it is a whitelisted constructor
\item type-casts with variable arguments (e.g., \mbox{\texttt{set(cache)}})
\item unpack operators (\mbox{\texttt{*args}} / \mbox{\texttt{**kwargs}})
\end{itemize}

\subsection{Static Analysis vs. LLM Fallback Boundary}
\label{app:dvp-boundary}

DVP has a strict boundary between static parsing and fallback.
\begin{itemize}[leftmargin=*,noitemsep,topsep=2pt]
\item If programmatic parsing cannot find a valid mask position that passes both funnels, the sample is discarded.
\item If the mask position is valid but the extracted answer is a variable name, fallback is allowed.
\end{itemize}

In fallback mode, the model acts only as a \emph{variable resolver}: it must unfold the variable into a concrete literal/constant/constructor/attribute form derived from the execution context. The resolved result is re-validated through both funnels.

\subsubsection{Variable Resolver Prompt}
\label{app:dvp-prompt-resolver}

When fallback is triggered, the LLM receives a prompt emphasizing runtime behavior reasoning:

\noindent\textbf{Key Requirements.}
\begin{itemize}[leftmargin=*,noitemsep,topsep=2pt]
\item \mbox{\textbf{Not code completion}}: Task is to infer runtime behavior based on trace information
\item \mbox{\textbf{Answer format}}: Must be a specific constant value or variable name (not complex expressions)
\item \mbox{\textbf{String quoting}}: String values must be wrapped in double quotes in JSON
\item \mbox{\textbf{Identity \textit{vs.} equality}}: Distinguish \mbox{\texttt{is}} (identity) from \mbox{\texttt{==}} (equality)
\item \mbox{\textbf{Hint quality}}: Must guide reasoning without leaking the answer
\end{itemize}

\noindent\textbf{Prompt Structure.}
The prompt includes:
\begin{itemize}[leftmargin=*,noitemsep,topsep=2pt]
\item Complete assert statement with context
\item Full test function code
\item Execution flow (trace information with function calls)
\item Input parameters
\item Design requirements emphasizing runtime behavior reasoning
\item Answer requirements specifying format constraints
\end{itemize}

\subsection{End-to-End Pseudocode}
\label{app:dvp-pseudo}
\begin{dvpbox}[title={DVP Algorithm}]

Algorithm~\ref{alg:dvp} summarizes the end-to-end Deterministic Value Protocol.

\end{dvpbox}

\begin{algorithm}[t]
\caption{Deterministic Value Protocol (DVP)}
\label{alg:dvp}
\footnotesize
\begin{algorithmic}[1]
\Require Trace record $R$, source code $S$
\Ensure Deterministic fill-in-the-blank set
\Statex \hspace{2em}$\mathcal{Q}$
\algstage{Trace parsing}
\State $T \gets \textsc{ExtractTraceInfo}(R)$
\State $A \gets \textsc{ParseAssertions}(S)$
\State $\mathcal{Q} \gets \emptyset$
\algstage{Candidate evaluation}
\For{each assertion $a$ in $A$}
  \State $c \gets \textsc{MaskCandidate}(a, T)$
  \If{$c = \bot$}
    \State \textbf{continue} \Comment{No deterministic mask}
  \EndIf
  \If{$\neg\, \textsc{SemanticDeterminism}(c, a, T)$}
    \State \textbf{continue} \Comment{Fails Funnel 1}
  \EndIf
  \If{$\textsc{MorphologyPriority}(c, T) = 0$}
    \State \textbf{continue} \Comment{Fails Funnel 2}
  \EndIf
  \If{$\textsc{IsVariable}(c)$ \textbf{and} $\textsc{FallbackAllowed}$}
    \algstage{Fallback resolution}
    \State $c' \gets \textsc{ResolveVariable}(c, T)$
    \If{$\neg\, \textsc{SemanticDeterminism}(c', a, T)$}
      \State \textbf{continue}
    \EndIf
    \If{$\textsc{MorphologyPriority}(c', T) = 0$}
      \State \textbf{continue}
    \EndIf
    \State $c \gets c'$
  \EndIf
  \State $\mathcal{Q} \gets \mathcal{Q} \cup$
  \Statex \hspace{2em}$\{\textsc{EmitQuestion}(a, c)\}$
\EndFor
\algstage{Batch emission}
\State \Return $\mathcal{Q}$
\end{algorithmic}
\end{algorithm}

\subsection{Concrete Example from Real Data}
\label{app:dvp-example}

Funnel~1 rejects unstable object repr strings. For example, in a representative dataset we observe inputs containing:
\begin{lstlisting}
<tests.test_dunders.TestEqOrder object at 0x7f6a9d5d38f0>
\end{lstlisting}
DVP rejects such answers, ensuring the final ground truth is stable under strict comparison.

Funnel~2 rejects non-explicit truths such as pure variable names (e.g., \texttt{result}) and dynamic calls (e.g., \texttt{func(value)}), which would otherwise permit shallow syntactic splicing without reasoning about runtime state.

\section{Cognitive Metrics Computation}
\label{app:trace-metrics}

This section discloses the inputs, assumptions, derivations, and outputs behind the Cognitive Metrics so that any researcher can reproduce ESV/MCL/DFI in an isolated environment. The main paper merely consumes the reported scores; implementation details, hyperparameters, algorithm descriptions, and samples live in this appendix.

\subsection{Artifacts and Input Contracts}

\noindent\textbf{Trace Record.}
Each JSONL entry represents a single test execution with \texttt{function\_calls} (array of \texttt{call\_order}, \texttt{function\_name}, \texttt{file\_path}, \texttt{function\_source\_code}, \texttt{executed\_lines}), a condensed \texttt{execution\_flow}, and assertion metadata (\texttt{assert\_line} counted from the dedented body, \texttt{line\_offset} for the absolute first line).

\noindent\textbf{Collector Configuration.}
All experiments fix \texttt{max\_trace\_size=1000}, \texttt{max\_call\_depth=3}, and \texttt{include\_test\_files=True}, with optional \texttt{target\_modules} to bound observation. The configuration is embedded in \texttt{input\_data} together with the \texttt{test\_function} identifier.

\noindent\textbf{Output Contract.}
The metric calculator consumes a single trace and emits \texttt{esv}, \texttt{mcl}, \texttt{dfi}, plus diagnostic sets such as \texttt{relevant\_calls}, \texttt{relevant\_lines}, and \texttt{source\_variables}, enabling verification without rerunning the slice.

\subsection{Interprocedural Slice and Evidence Harvesting}

The implementation transforms a single trace into slice evidence via a strict pipeline so that every metric consumes the same structured input.

\begin{enumerate}[leftmargin=*,noitemsep,topsep=2pt]
  \item \mbox{\textbf{Trace Normalization}}: Build a \texttt{call\_index} and \texttt{line\_to\_call} map over \texttt{function\_calls} so each absolute line immediately resolves to its \texttt{call\_order}. Empty frames are discarded and the entry frame is cached as \texttt{entry\_call} to anchor later steps.
  \item \mbox{\textbf{Assertion Anchor Extraction}}: \textit{\_extract\_assert\_targets\_with\_beniget} dedents the entry source, uses beniget def-use chains to enumerate every \texttt{gast.Name} read on the assertion line, and captures the call expressions that appear in the same line. A regex fallback is used only if the analysis fails, yielding both the target variable set and the function calls triggered\allowbreak{} by the assertion.
  \item \mbox{\textbf{Worklist Seeding}}: Targets are enqueued as $(0, \text{targets}, \text{"assert\_vars"})$, while assertion-triggered callees are enqueued as $(\text{call\_order}, \{\text{"\_\_return\_\_"}\}, \text{"called\_from\_*"})$ so their returns are tracked through a synthetic placeholder. Decorator/dynamic heuristics are evaluated before seeding to attach \texttt{original\_name} and \texttt{is\_decorator} flags,\allowbreak{} preventing double counting.
  \item \mbox{\textbf{Frame-level Resolution}}: The LIFO worklist pops $(\text{call\_order}, \text{target\_vars}, \text{context})$, retrieves \texttt{call\_info}, converts relative assertion lines to absolutes (entry uses \texttt{line\_offset}, other frames use their own \texttt{line\_number}), and invokes \textit{\_slice\_within\_function} to obtain \texttt{relevant\_lines}, \texttt{mutations}, \texttt{source\_vars}, and \texttt{function\_calls}. Execution counts are accumulated into the global \texttt{relevant\_lines} map.
  \item \mbox{\textbf{Dependency Expansion}}: Function-level \texttt{function\_calls} are matched against the \texttt{call\_index}; matched frames are re-enqueued with $\{\text{"\_\_return\_\_"}\}$ targets, while missing matches are captured in \texttt{untraced\_function\_calls} to expose trace coverage holes.
\end{enumerate}

The resulting \texttt{relevant\_calls}, \texttt{relevant\_lines}, and \texttt{source\_variables} form the unique evidence set for all metrics. Section~\ref{app:trace-metrics-worklist} details the fine-grained \textit{\_slice\_within\_function} solver, and Section~\ref{app:trace-metrics-defs} derives ESV/MCL/DFI from that evidence.

\subsection{Worklist Resolution Strategy}
\label{app:trace-metrics-worklist}

\textit{calculate\_interprocedural\_slice} drives \textit{\_slice\_within\_function} with a strict LIFO worklist so that each frame undergoes AST-level backward slicing.

\noindent\textbf{Per-frame pipeline.}
\begin{enumerate}[leftmargin=*,noitemsep,topsep=2pt]
  \item \mbox{\textbf{Source Canonicalization}}: Fetch \texttt{function\_source\_code}, dedent it via \texttt{textwrap.dedent}, and build \texttt{abs\_to\_rel}/\texttt{rel\_to\_abs} maps so entry and nested frames share one coordinate system.
  \item \mbox{\textbf{Executed-line Alignment}}: Map the recorded absolute lines to the relative set \texttt{executed\_rel}, ensuring non-executed statements never enter the slice.
  \item \mbox{\textbf{Def-use and Control Dependencies}}: Invoke \textit{ExecutionFlowParser.\_build\_beniget\_chains} to build def-use chains and run \textit{ControlDependencyAnalyzer} to map \texttt{line\_no} $\to$ \textit{controlling lines}, feeding explicit control edges into the backward pass.
  \item \mbox{\textbf{Call Extraction}}: Walk the dedented AST to capture every \texttt{gast.Call} within \texttt{executed\_rel}, storing its line number, callee name, and argument variables for cross-frame propagation.
  \item \mbox{\textbf{Backward Expansion}}: Initialize \texttt{needed\_vars} = \texttt{target\_vars} and keep popping variables. For each definition found in \texttt{variable\_writes}, call \texttt{add\_relevant\_line} to add it into \texttt{relevant\_lines\_rel}, close over control dependencies, and enqueue any variables read on that line. Entry frames under \textit{context="assert\_vars"} force the assertion line plus all of its controllers to stay in the slice, whereas return contexts focus on \texttt{return} statements.
  \item \mbox{\textbf{Source-variable Inference}}: After \texttt{relevant\_lines\_rel} is determined, count the variables that were read but never written within that set, subtracting builtins, keywords, imported symbols, and detected callee names. \textit{\_analyze\_semantic\_dependencies} inspects complex expressions (e.g., dictionaries or higher-order calls) to append their true data dependencies into \texttt{source\_vars}.
\end{enumerate}

\noindent\textbf{Propagation semantics.}
\begin{itemize}[leftmargin=*,noitemsep,topsep=2pt]
  \item Each state key \textit{(call\_order, frozenset(target\_vars), context)} ensures a frame/target pair executes at most once even under deep recursion or decorators.
  \item Returned \texttt{function\_calls} are matched against \texttt{call\_index}; hits are re-enqueued with \textit{\{"\_\_return\_\_"\}} targets, while misses get logged in \textit{untraced\_function\_calls} to expose coverage gaps.
\end{itemize}

Once the worklist drains, \texttt{relevant\_calls}, \texttt{relevant\_lines}, and \texttt{source\_variables} are fixed, enabling direct metric computation without further heuristics.

\subsection{Metric Definitions and Fixed Assumptions}
\label{app:trace-metrics-defs}

Every metric uses the same interprocedural slice. Let $\mathcal{F}_{\text{rel}}$ denote the relevant functions, $\mathcal{L}_{\text{rel}}$ the multiset of relevant lines with execution counts $c$, and $\mathcal{S}$ the variables read but not reassigned inside the slice.

\begin{align*}
\text{ESV} &= \sum_{f \in \mathcal{F}_{\text{rel}}} \text{LOC}(f) \\
\text{MCL} &= \sum_{(\ell, c) \in \mathcal{L}_{\text{rel}}} c \\
\text{DFI} &= |\mathcal{S}|
\end{align*}

\begin{itemize}[leftmargin=*,noitemsep,topsep=2pt]
  \item \mbox{\textbf{ESV (Effective Sliced Volume)}}: Deduplicate and count lines of code across $\mathcal{F}_{\text{rel}}$ to capture the unique reading load needed to justify the assertion.
  \item \mbox{\textbf{MCL (Mutation Chain Length)}}: Sum execution frequencies in $\mathcal{L}_{\text{rel}}$ to characterize the dynamic simulation depth.
  \item \mbox{\textbf{DFI (Dependency Fan-In)}}: Count elements in $\mathcal{S}$ to describe how many independent information sources feed the assertion.
\end{itemize}

\subsubsection{ESV Computation Steps}
\begin{enumerate}[leftmargin=*,noitemsep,topsep=2pt]
  \item Gather deduplicated source snippets for every function in $\mathcal{F}_{\text{rel}}$, counting executable lines only (whitespace and comments are skipped).
  \item Record each function once even if it is invoked multiple times, preventing duplicate reading cost.
  \item Summing all unique functions' line counts yields ESV, the minimal code volume a reviewer must inspect.
\end{enumerate}

\subsubsection{MCL Computation Steps}
\begin{enumerate}[leftmargin=*,noitemsep,topsep=2pt]
  \item Maintain a map from each statement in $\mathcal{L}_{\text{rel}}$ to its execution count derived from \texttt{executed\_lines}.
  \item Sum the counts to obtain MCL; hot paths where the same line fires repeatedly increase the metric proportionally.
  \item Identical source lines in distinct frames accumulate separately so that MCL reflects the actual dynamic chain length.
\end{enumerate}

\subsubsection{DFI Computation Steps}
\begin{enumerate}[leftmargin=*,noitemsep,topsep=2pt]
  \item During dependency propagation, collect every variable that is read but never reassigned, forming the set $\mathcal{S}$.
  \item Deduplicate the set so that a variable counts once regardless of how many frames reference it.
  \item The cardinality of $\mathcal{S}$ equals DFI, indicating how many independent information sources must be tracked; higher values imply heavier contextual load.
\end{enumerate}

The calculator exposes no tunable hyperparameters; interprocedural slicing is always enabled so that all three scores rely on identical evidence.

\subsection{Representative Example}

We choose the real trace \texttt{question\_id=544\_3} from the toolz corpus, anchored at the function \texttt{test\_accumulate}. The test chains three representative scenarios—seedless accumulation, seeded accumulation, and empty-sequence fallbacks—so that every major argument pattern of \texttt{accumulate} is exercised.

\noindent\textbf{Test body excerpt.}
\begin{lstlisting}[style=tracefig]
def test_accumulate():
    numbers = [2, 1, 4, 3, 7]
    assert list(accumulate(add, numbers)) == [2, 3, 7, 10, 17]
\end{lstlisting}

\noindent\textbf{accumulate function implementation.}
\begin{lstlisting}[style=tracefig]
def accumulate(binop, seq, initial=no_default):
    """ Repeatedly apply binary function to a sequence, accumulating results

    >>> from operator import add, mul
    >>> list(accumulate(add, [1, 2, 3, 4, 5]))
    [1, 3, 6, 10, 15]
    >>> list(accumulate(mul, [1, 2, 3, 4, 5]))
    [1, 2, 6, 24, 120]

    Accumulate is similar to ``reduce`` and is good for making functions like
    cumulative sum:

    >>> from functools import partial, reduce
    >>> sum    = partial(reduce, add)
    >>> cumsum = partial(accumulate, add)

    Accumulate also takes an optional argument that will be used as the first
    value. This is similar to reduce.

    >>> list(accumulate(add, [1, 2, 3], -1))
    [-1, 0, 2, 5]
    >>> list(accumulate(add, [], 1))
    [1]

    See Also:
        itertools.accumulate :  In standard itertools for Python 3.2+
    """
    seq = iter(seq)
    if initial == no_default:
        try:
            result = next(seq)
        except StopIteration:
            return
    else:
        result = initial
    yield result
    for elem in seq:
        result = binop(result, elem)
        yield result
\end{lstlisting}

\textit{calculate\_interprocedural\_slice} binds the entry frame and the \texttt{accumulate} implementation into one evidence pool: the entry frame supplies the sequences and assertions, while the library frame expands every intermediate state.

\begin{table}[t]
\scriptsize
\centering
\caption{Metric values for the \texttt{test\_accumulate} example.}
\label{tab:trace-example}
\setlength{\tabcolsep}{1.5pt}
\begin{tabular}{p{0.18\columnwidth} p{0.11\columnwidth} p{0.65\columnwidth}}
\toprule
\textbf{Metric} & \textbf{Value} & \textbf{Evidence Source} \\
\midrule
ESV & 59 lines & $\mathcal{F}_{\text{rel}} = \{\text{test\_accumulate}, \text{accumulate}\}$; deduplicated source from the entry test plus the library implementation totals 59 executable lines,\allowbreak{} mirroring the minimal reading load. \\
MCL & 35 & $\mathcal{L}_{\text{rel}}$ spans the entry assertions plus 11 relevant lines inside \texttt{accumulate}; the trace repeats those lines across the accumulate variants,\allowbreak{} yielding 35 total executions. \\
DFI & 3 & The slice retains three read-only dependencies—\texttt{add}, \texttt{initial}, and \texttt{no\_default}—all supplied via imports or default guards,\allowbreak{} hence $|\mathcal{S}| = 3$. \\
\bottomrule
\end{tabular}
\end{table}

\noindent\textbf{Metric derivations.}
\begin{itemize}[leftmargin=*,noitemsep,topsep=2pt]
  \item \mbox{\textbf{ESV}}: Understanding how the folds evolve under different seeds requires reviewing both the test harness and \texttt{accumulate}, hence their combined 59 lines define ESV.
  \item \mbox{\textbf{MCL}}: $\mathcal{L}_{\text{rel}}$ covers the entry assertions and 11 relevant lines inside \texttt{accumulate}; the trace shows these lines are repeatedly accessed across multiple accumulation branches,\allowbreak{} with execution counts summing to 35.
  \item \mbox{\textbf{DFI}}: During propagation, the read-only dependencies expand to three symbols—\texttt{add}, \texttt{initial}, and \texttt{no\_default}—each from module imports or default parameter constraints,\allowbreak{} so $|\mathcal{S}| = 3$.
\end{itemize}

This example shows how a single interprocedural slice lets us reason about multiple assertion branches: even when the assertions live entirely in the entry frame, the slicing framework still brings in the library implementation to keep metric computation reproducible.

\section{DFI Case Studies: Integration Failure in Practice}
\label{app:dfi_cases}

To illustrate what Integration Width (DFI) looks like in practice and how it induces Aggregation Deficits in agents, we present two concrete case studies. DFI ($\text{DFI} = |\mathcal{S}|$) measures the number of independent external variables in a dynamic slice that are read but never reassigned, reflecting how many disparate information sources an agent must simultaneously integrate to deduce the verified state.

\subsection{Case 1: Jinja2 \texttt{test\_groupby\_default} (DFI=16)}

\begin{lstlisting}[language=Python]
def test_groupby_default(self, env):
    template_src = "".join([
        "{% for grp, bucket in people|groupby('city', default='SF') %}",
        "{{ grp }}: {{ bucket|map(attribute='name')|join(' | ') }}\n",
        "{% endfor %}",
    ])
    tmpl = env.from_string(template_src)
    payload = [
        {"name": "alex", "city": "SF"},
        {"name": "brian", "city": "WA"},
        {"name": "cora"},                    # <-- no city key
        {"name": "dana", "city": "SF"},
        {"name": "erin", "city": "WA"},
    ]
    out = tmpl.render(people=payload)
    assert out == ___  # correct: "SF: alex | cora | dana\nWA: brian | erin\n"
\end{lstlisting}

The 16 source variables span four context layers: template compilation (\texttt{code}, \texttt{source}, etc.), runtime data (\texttt{args}, \texttt{kwargs}), filter chain (\texttt{value}), and environment configuration (\texttt{env}, \texttt{environment}, and others). The agent produced \texttt{"SF: alex | dana\textbackslash nWA: brian | erin\textbackslash nSF: cora\textbackslash n"} (incorrect).

\textbf{Failure Mechanism}: The agent correctly inferred the \texttt{default='SF'} semantics (tagging \texttt{cora} as \texttt{SF}) but produced two disjoint \texttt{SF} groups. Jinja2's \texttt{groupby} filter sorts elements before grouping (\texttt{sorted(value, key=expr)}), which merges all \texttt{SF} entries. Although the agent's trajectory showed it read the documentation for \texttt{sync\_do\_groupby} specifying this sorting behavior, it failed to integrate this third logical layer. The aggregation chain broke after successfully integrating the first two layers.

\subsection{Case 2: NetworkX \texttt{test\_all\_simple\_paths} (DFI=10)}

\begin{lstlisting}[language=Python]
def test_all_simple_paths_on_non_trivial_graph():
    H = nx.path_graph(6, create_using=nx.DiGraph())
    extra_arcs = [(0, 6), (2, 6), (2, 4), (6, 5), (5, 3), (5, 4)]
    H.add_edges_from(extra_arcs)

    route_iter = nx.all_simple_paths(H, 2, [3, 4], cutoff=2)
    observed = {tuple(step_list) for step_list in route_iter}
    assert observed == ___  # correct: {(2, 3), (2, 4), (2, 3, 4)}
\end{lstlisting}

The \texttt{path\_graph(6)} constructs a 0$\to$1$\to$\dots$\to$5 chain, augmented by extra arcs. The \texttt{cutoff=2} parameter restricts valid paths to those with $\le 2$ edges. The 10 source variables span graph construction, path enumeration logic, and the truncation constraint (\texttt{cutoff}). The agent predicted \texttt{\{(2,3), (2,4), (2,3,4), (2,4,5,3), (2,6,5,3), (2,6,5,4), (2,6,5,3,4)\}} (incorrect).

\textbf{Failure Mechanism}: The agent correctly built the graph topology and enumerated paths but failed to integrate the \texttt{cutoff=2} constraint, returning paths with 3 and 4 edges. This exemplifies partial integration followed by breakage: as DFI increases, agents correctly process initial information layers but increasingly drop final constraints.

\section{Impact of Trace Depth Limit}

During dynamic trace extraction, we apply a maximum call depth limit (\texttt{max\_call\_depth=3}) to optimize the signal-to-noise ratio. To quantify its impact, we compared traces generated with \texttt{depth=3} versus unbounded \texttt{depth=$\infty$} across all repositories (Table~\ref{tab:trace_depth}).

\begin{table}[htbp]
\scriptsize
\centering
\renewcommand{\arraystretch}{1.35}
\caption{Trace Inflation Factor (\texttt{depth=3} vs.\ \texttt{depth=$\infty$})}
\label{tab:trace_depth}
\setlength{\tabcolsep}{1.8pt}
\begin{tabularx}{\columnwidth}{@{} l *{5}{>{\centering\arraybackslash}X} @{}}
\toprule
\textbf{Repo} & \makecell{\textbf{Sam-}\\\textbf{ples}} & \makecell{\textbf{Depth}\\\textbf{3}} & \makecell{\textbf{Depth}\\\textbf{$\infty$}} & \makecell{\textbf{Infla-}\\\textbf{tion}} & \makecell{\textbf{Max}\\\textbf{Dep.}} \\
\midrule
\rowcolor[HTML]{F5F5F5}
\texttt{cachetools} & 51  & 369.7 & 396.7  & 1.1x  & 11 \\
\texttt{toolz}      & 100 & 40.1  & 50.1   & 1.3x  & 7 \\
\rowcolor[HTML]{F5F5F5}
\texttt{yarl}       & 100 & 4.9   & 9.0    & 1.8x  & 10 \\
\texttt{networkx}   & 100 & 236.3 & 350.8  & 1.5x  & 14 \\
\rowcolor[HTML]{FFFDE7}
\texttt{jinja2}     & 100 & 20.4  & 981.4  & \textbf{48.0x} & 49 \\
\rowcolor[HTML]{F5F5F5}
\texttt{attrs}      & 100 & 9.1   & 11.6   & 1.3x  & 8 \\
\rowcolor[HTML]{FFFDE7}
\texttt{sympy}      & 100 & 272.1 & 3186.5 & \textbf{11.7x} & 36 \\
\midrule
\rowcolor[HTML]{E8F5E9}
\textbf{Overall}    & \textbf{651} & \textbf{118.5} & \textbf{736.1} & \textbf{6.2x} & \textbf{--} \\
\bottomrule
\end{tabularx}
\end{table}

Removing the depth limit inflates function call counts by 6.2x overall. The vast majority of added calls correspond to low-level infrastructure functions (AST traversers, iterator protocols, magic methods) that do not contribute to the high-level semantic understanding of the logic. Crucially, this depth limit applies \emph{only} to trace extraction for metric computation; during reasoning, the evaluating agent retains unconstrained access to the complete repository codebase and execution environment.

\section{Difficulty Classification Protocol}
\label{app:difficulty}

To control for task variance, instance difficulty is explicitly stratified based on AI reasoning complexity. This annotation is generated by an independent evaluator LLM (GPT-5.2) during the data generation phase. The LLM acts solely as an annotator, analyzing the blanked assertion and execution trace to assign a label of \texttt{easy}, \texttt{medium}, or \texttt{hard}. The prompt template is provided in Box~\ref{box:difficulty}.

\begin{agentbox}[Difficulty Classification Prompt Template]
\label{box:difficulty}
\small
You are an expert at evaluating programming question difficulty for AI systems.

\textbf{Given:}
\begin{itemize}
    \item \textbf{QUESTION:} (the blanked assertion)
    \item \textbf{CONTEXT:} (execution-trace context for understanding only; do NOT assume the answer is visible)
\end{itemize}

\textbf{Task:}
Classify the instance difficulty into one of \{easy, medium, hard\} based on AI reasoning complexity. Use the following criteria:

\begin{itemize}
    \item \textbf{easy}: Requires only basic Python knowledge and direct code reading. Involves linear logic, shallow abstraction, and minimal state tracking.
    \item \textbf{medium}: Requires understanding non-trivial API behavior, simple state mutations, control flow loops, or basic algorithms.
    \item \textbf{hard}: Requires deep domain knowledge (e.g., symbolic math, complex graph theory), resolving deep multi-layer dependencies, extensive state tracking, or metaprogramming.
\end{itemize}

Provide your classification exactly as one of the three keywords.
\end{agentbox}

\end{document}